\documentclass[preprint]{aastex701}

\usepackage{amsmath}
\usepackage{booktabs}
\usepackage{hyperref}
\begin{document}

\title{Time-Integrated Searches for Sub-TeV Neutrino Sources with IceCube-DeepCore}

\email[show]{\href{mailto:analysis@icecube.wisc.edu}{analysis@icecube.wisc.edu}}
\affiliation{III. Physikalisches Institut, RWTH Aachen University, D-52056 Aachen, Germany}
\affiliation{Department of Physics, University of Adelaide, Adelaide, 5005, Australia}
\affiliation{Dept. of Physics and Astronomy, University of Alaska Anchorage, 3211 Providence Dr., Anchorage, AK 99508, USA}
\affiliation{School of Physics and Center for Relativistic Astrophysics, Georgia Institute of Technology, Atlanta, GA 30332, USA}
\affiliation{Dept. of Physics, Southern University, Baton Rouge, LA 70813, USA}
\affiliation{Dept. of Physics, University of California, Berkeley, CA 94720, USA}
\affiliation{Lawrence Berkeley National Laboratory, Berkeley, CA 94720, USA}
\affiliation{Institut f{\"u}r Physik, Humboldt-Universit{\"a}t zu Berlin, D-12489 Berlin, Germany}
\affiliation{Fakult{\"a}t f{\"u}r Physik {\&} Astronomie, Ruhr-Universit{\"a}t Bochum, D-44780 Bochum, Germany}
\affiliation{Universit{\'e} Libre de Bruxelles, Science Faculty CP230, B-1050 Brussels, Belgium}
\affiliation{Vrije Universiteit Brussel (VUB), Dienst ELEM, B-1050 Brussels, Belgium}
\affiliation{Dept. of Physics, Simon Fraser University, Burnaby, BC V5A 1S6, Canada}
\affiliation{Department of Physics and Laboratory for Particle Physics and Cosmology, Harvard University, Cambridge, MA 02138, USA}
\affiliation{Dept. of Physics, Massachusetts Institute of Technology, Cambridge, MA 02139, USA}
\affiliation{Dept. of Physics and The International Center for Hadron Astrophysics, Chiba University, Chiba 263-8522, Japan}
\affiliation{Department of Physics, Loyola University Chicago, Chicago, IL 60660, USA}
\affiliation{Dept. of Physics and Astronomy, University of Canterbury, Private Bag 4800, Christchurch, New Zealand}
\affiliation{Dept. of Physics, University of Maryland, College Park, MD 20742, USA}
\affiliation{Dept. of Astronomy, Ohio State University, Columbus, OH 43210, USA}
\affiliation{Dept. of Physics and Center for Cosmology and Astro-Particle Physics, Ohio State University, Columbus, OH 43210, USA}
\affiliation{Niels Bohr Institute, University of Copenhagen, DK-2100 Copenhagen, Denmark}
\affiliation{Dept. of Physics, TU Dortmund University, D-44221 Dortmund, Germany}
\affiliation{Dept. of Physics and Astronomy, Michigan State University, East Lansing, MI 48824, USA}
\affiliation{Dept. of Physics, University of Alberta, Edmonton, Alberta, T6G 2E1, Canada}
\affiliation{Erlangen Centre for Astroparticle Physics, Friedrich-Alexander-Universit{\"a}t Erlangen-N{\"u}rnberg, D-91058 Erlangen, Germany}
\affiliation{Physik-department, Technische Universit{\"a}t M{\"u}nchen, D-85748 Garching, Germany}
\affiliation{D{\'e}partement de physique nucl{\'e}aire et corpusculaire, Universit{\'e} de Gen{\`e}ve, CH-1211 Gen{\`e}ve, Switzerland}
\affiliation{Dept. of Physics and Astronomy, University of Gent, B-9000 Gent, Belgium}
\affiliation{Dept. of Physics and Astronomy, University of California, Irvine, CA 92697, USA}
\affiliation{Karlsruhe Institute of Technology, Institute for Astroparticle Physics, D-76021 Karlsruhe, Germany}
\affiliation{Karlsruhe Institute of Technology, Institute of Experimental Particle Physics, D-76021 Karlsruhe, Germany}
\affiliation{Dept. of Physics, Engineering Physics, and Astronomy, Queen's University, Kingston, ON K7L 3N6, Canada}
\affiliation{Department of Physics {\&} Astronomy, University of Nevada, Las Vegas, NV 89154, USA}
\affiliation{Nevada Center for Astrophysics, University of Nevada, Las Vegas, NV 89154, USA}
\affiliation{Dept. of Physics and Astronomy, University of Kansas, Lawrence, KS 66045, USA}
\affiliation{UCLouvain, Centre for Cosmology, Particle Physics and Phenomenology, CP3, Chemin du Cyclotron 2, 1348 Louvain-la-Neuve, Belgium}
\affiliation{Department of Physics, Mercer University, Macon, GA 31207-0001, USA}
\affiliation{Dept. of Astronomy, University of Wisconsin{\textemdash}Madison, Madison, WI 53706, USA}
\affiliation{Dept. of Physics and Wisconsin IceCube Particle Astrophysics Center, University of Wisconsin{\textemdash}Madison, Madison, WI 53706, USA}
\affiliation{Institute of Physics, University of Mainz, Staudinger Weg 7, D-55099 Mainz, Germany}
\affiliation{Department of Physics, Marquette University, Milwaukee, WI 53201, USA}
\affiliation{Institut f{\"u}r Kernphysik, Universit{\"a}t M{\"u}nster, D-48149 M{\"u}nster, Germany}
\affiliation{Bartol Research Institute and Dept. of Physics and Astronomy, University of Delaware, Newark, DE 19716, USA}
\affiliation{Dept. of Physics, Yale University, New Haven, CT 06520, USA}
\affiliation{Columbia Astrophysics and Nevis Laboratories, Columbia University, New York, NY 10027, USA}
\affiliation{Dept. of Physics, University of Oxford, Parks Road, Oxford OX1 3PU, United Kingdom}
\affiliation{Dipartimento di Fisica e Astronomia Galileo Galilei, Universit{\`a} Degli Studi di Padova, I-35122 Padova PD, Italy}
\affiliation{Dept. of Physics, Drexel University, 3141 Chestnut Street, Philadelphia, PA 19104, USA}
\affiliation{Physics Department, South Dakota School of Mines and Technology, Rapid City, SD 57701, USA}
\affiliation{Dept. of Physics, University of Wisconsin, River Falls, WI 54022, USA}
\affiliation{Dept. of Physics and Astronomy, University of Rochester, Rochester, NY 14627, USA}
\affiliation{Department of Physics and Astronomy, University of Utah, Salt Lake City, UT 84112, USA}
\affiliation{Dept. of Physics, Chung-Ang University, Seoul 06974, Republic of Korea}
\affiliation{Oskar Klein Centre and Dept. of Physics, Stockholm University, SE-10691 Stockholm, Sweden}
\affiliation{Dept. of Physics and Astronomy, Stony Brook University, Stony Brook, NY 11794-3800, USA}
\affiliation{Dept. of Physics, Sungkyunkwan University, Suwon 16419, Republic of Korea}
\affiliation{Institute of Physics, Academia Sinica, Taipei, 11529, Taiwan}
\affiliation{Dept. of Physics and Astronomy, University of Alabama, Tuscaloosa, AL 35487, USA}
\affiliation{Dept. of Astronomy and Astrophysics, Pennsylvania State University, University Park, PA 16802, USA}
\affiliation{Dept. of Physics, Pennsylvania State University, University Park, PA 16802, USA}
\affiliation{Dept. of Physics and Astronomy, Uppsala University, Box 516, SE-75120 Uppsala, Sweden}
\affiliation{Dept. of Physics, University of Wuppertal, D-42119 Wuppertal, Germany}
\affiliation{Deutsches Elektronen-Synchrotron DESY, Platanenallee 6, D-15738 Zeuthen, Germany}

\author[0000-0001-6141-4205]{R. Abbasi}
\affiliation{Department of Physics, Loyola University Chicago, Chicago, IL 60660, USA}
\email{rabbasi@luc.edu}

\author[0000-0001-8952-588X]{M. Ackermann}
\affiliation{Deutsches Elektronen-Synchrotron DESY, Platanenallee 6, D-15738 Zeuthen, Germany}
\email{markus.ackermann@desy.de}

\author{J. Adams}
\affiliation{Dept. of Physics and Astronomy, University of Canterbury, Private Bag 4800, Christchurch, New Zealand}
\email{jenni.adams@canterbury.ac.nz}

\author[0000-0003-2252-9514]{J. A. Aguilar}
\affiliation{Universit{\'e} Libre de Bruxelles, Science Faculty CP230, B-1050 Brussels, Belgium}
\email{juanan.aguilar@icecube.wisc.edu}

\author[0000-0003-0709-5631]{M. Ahlers}
\affiliation{Niels Bohr Institute, University of Copenhagen, DK-2100 Copenhagen, Denmark}
\email{mahlers@icecube.wisc.edu}

\author[0000-0002-9534-9189]{J.M. Alameddine}
\affiliation{Dept. of Physics, TU Dortmund University, D-44221 Dortmund, Germany}
\email{jean-marco.alameddine@icecube.wisc.edu}

\author[0009-0001-2444-4162]{S. Ali}
\affiliation{Dept. of Physics and Astronomy, University of Kansas, Lawrence, KS 66045, USA}
\email{shoukat@ku.edu}

\author{N. M. Amin}
\affiliation{Bartol Research Institute and Dept. of Physics and Astronomy, University of Delaware, Newark, DE 19716, USA}
\email{moureen@udel.edu}

\author[0000-0001-9394-0007]{K. Andeen}
\affiliation{Department of Physics, Marquette University, Milwaukee, WI 53201, USA}
\email{karen.andeen@icecube.wisc.edu}

\author[0000-0003-4186-4182]{C. Arg{\"u}elles}
\affiliation{Department of Physics and Laboratory for Particle Physics and Cosmology, Harvard University, Cambridge, MA 02138, USA}
\email{carlos.arguelles@icecube.wisc.edu}

% ...
\date{\today}
\collaboration{432}{The IceCube Collaboration}


\begin{abstract}

We have developed techniques for a competitive sub-TeV time-integrated neutrino search and applied it to 11.1 years of IceCube-DeepCore data. The DeepCore subarray lowers the sensitivity of IceCube down to sub-TeV energies and is especially interesting for objects with soft spectra. Three studies were performed: a search for neutrino emission from AGN exhibiting high intrinsic X-ray flux, including NGC 1068, as identified by SWIFT/BAT; a search for neutrino emission from Galactic objects identified by Fermi-LAT as exhibiting a spectral shape consistent with neutral pion decay; and an all-sky search for neutrino point sources. Objects for this study were selected given their prospects for sub-TeV neutrino emission. No evidence for sub-TeV neutrino emission is found in any of the searches performed. Finally, for each catalog of objects, we use a statistical combination of the p-values via a binomial test to search for aggregated neutrino emission from a subset of the objects. Neither of the binomial tests yields significant results. For NGC 1068, assuming a power law spectrum with index 3.4, the 90\% confidence level upper limit on per-flavor neutrino emission in the 30--400~GeV range is $\Phi_{\nu+\bar{\nu}}|_{\mathrm{1 TeV}} < 9.5 \times 10^{-11}$ TeV$^{-1}$ cm$^{-2}$ s$^{-1}$, a factor of two higher than the extrapolation of IceCube's measurement at higher energies. We additionally provide neutrino flux upper limits for a variety of spectra.

\end{abstract}



\section{Introduction} \label{sec:intro}

The IceCube Neutrino Observatory \citep{Icecube:2016Detector} has opened the field of high-energy neutrino astrophysics through a series of results: the detection of a diffuse all-sky neutrino flux \citep{IceCube:2013Science}; evidence of neutrino emission from the blazar TXS~0506+056 both in temporal coincidence with a $\gamma$-ray flare \citep{IceCube:2018ScienceAlert} and during a $\gamma$-ray quiet period \citep{IceCube:2018ScienceFlare}; evidence for neutrino emission from the Seyfert galaxy NGC~1068 \citep{IceCube:2022Science,abbasi2025evidenceneutrinoemissionxray}; and the detection of neutrinos from the Milky Way \citep{IceCube:2023Science}. Astrophysical neutrinos are intimately connected to cosmic rays and $\gamma$-rays through hadronic processes in which cosmic rays interact with matter and/or photon fields to produce charged and neutral pions. The subsequent decay of charged pions produces neutrinos, while the decay of neutral pions produces $\gamma$-rays. In some environments, $\gamma$-rays are efficiently absorbed whereas neutrinos escape unattenuated, making neutrinos unique probes of otherwise hidden hadronic accelerators \citep{Murase2016Hidden}. 

IceCube-DeepCore \citep{ABBASI2012615}, a subdetector of IceCube sensitive to sub-TeV energies, has been used in searches for $<$TeV transient astrophysical phenomena such as GRBs \citep{Abbasi2024Search}, including GRB 221009A \citep{Abbasi2023Limits}, novae \citep{Abbasi2023Search}, and gravitational wave events \citep{Abbasi2023SearchGW}. This work presents a time-integrated astrophysical study using DeepCore. We focus on soft-spectra sources with potentially large sub-TeV neutrino flux. We search for neutrinos from a high intrinsic X-ray flux galaxy catalog \citep{Ricci:2017ApJS}, including NGC 1068, a catalog of Galactic objects exhibiting a ``pion bump'' in GeV $\gamma$-rays measured by Fermi-LAT \citep{Abdollahi2022Search}, and an all-sky search. For each catalog, we also statistically combined the results to search for a subpopulation of sources that may be detectable only as an ensemble instead of as individual sources.

Our search for neutrinos from galaxies with high intrinsic X-ray fluxes is motivated by IceCube's measurement of NGC 1068 \citep{IceCube:2022Science}. We adopt the updated study by \citealt{abbasi2025evidenceneutrinoemissionxray} as a reference for the measurement of NGC~1068, which describes NGC~1068's neutrino spectrum as a power law with index $\gamma=3.4$ for neutrino energies between 0.3 and 3.9~TeV.

NGC 1068 has also been detected in GeV $\gamma$-rays \citep{1FGL}, likely dominated by star-formation activity \citep{Eichmann:2022ApJ}. The $\gamma$-ray flux is significantly lower than the neutrino flux, suggesting the obscured nucleus as the neutrino source \citep{Inoue:2019ApJ, Murase2023HighAGN}. We highlight two models for neutrino production in NGC 1068. \citealt{Murase:2020PhRvL} propose that protons are accelerated up to $\sim\,$PeV energies in the turbulent corona of the supermassive black hole (SMBH).  
UV/optical photons from the accretion disk are Compton up-scattered by electrons in the corona to $\sim\,$keV energies \citep{Padovani:2017AApR}. Accelerated protons interact with up-scattered photons, producing charged and neutral pions that decay into neutrinos and $\gamma$-rays, respectively. The $\gamma$-rays are absorbed in the vicinity of the SMBH. \citealt{Blanco2025Neutrino} propose protons accelerated in the corona interacts with gas, also resulting in pions. Both models for NGC 1068 depend on uncertain or unknown environmental conditions near the SMBH, including the strength of the coronal magnetic field, the size of the corona, and how turbulent the corona is. Both models demonstrate similar spectral behavior at $\gtrsim 2$~TeV, but sub-TeV measurements could help differentiate between them at hundreds of GeV.

We also analyze a catalog of Galactic objects identified by Fermi-LAT as exhibiting a ``pion-bump,'' or a sharp cut-off in the $\gamma$-ray spectrum below GeV energies, indicative of neutral pion decay \citep{Abdollahi2022Search}. These sources often demonstrate a soft $\gamma$-ray spectrum above 100~GeV \citep{4FGL}, favoring study with DeepCore. The neutrino emission of the Milky Way \citep{IceCube:2023Science} is likely due to a combination of diffuse emission from cosmic-ray propagation in the Galaxy and unresolved individual sources. However, the relative contribution between the diffuse emission and the individual sources is unclear. In this work, we search for neutrinos from individual Galactic sources, focusing on sub-TeV emission.

\section{IceCube, DeepCore, and Source Catalog Description} \label{sec:data}

\subsection{Detector and Dataset} \label{subsec:detdata}

Located at the South Pole, IceCube \citep{Icecube:2016Detector} constitutes a cubic kilometer array of 5,160 digital optical modules (DOMs) suspended on 86 strings buried between 1.45 and 2.45 kilometers below the surface. Each DOM is equipped with a photomultiplier tube (PMT) that detects Cherenkov light emitted by relativistic charged particles \citep{Abbasi2010Calibration}. The main IceCube array is best suited for studying $\geq$TeV-scale neutrinos, with diminishing sensitivity at lower energies. DeepCore \citep{ABBASI2012615} is located in the bottom and center of IceCube. In DeepCore, strings are installed more closely, DOMs have reduced vertical spacing, and a fraction of DOMs have higher quantum efficiency PMTs. This enables DeepCore to detect neutrinos down to a few GeV. Charge current $\nu_\mu$ interactions occurring within or near the detector result in a cascade plus an outgoing muon, with the latter leaving a long track of Cherenkov light in the detector. This track-like morphology constitutes the overwhelming majority of the data analyzed in this study.

For this study, we use a modified version of the GRECO-Astronomy (GeV Reconstructed Events with Containment for Oscillations) dataset. With 11.1 years of livetime from April 26, 2012 to November 28, 2023, this dataset has been used for oscillation \citep{Aartsen2019MeasurementNuTau} and astrophysical transient studies \citep{Abbasi2023Search}. However, performing time-integrated studies requires improvements in angular resolution, which this work addresses by introducing additional cuts and an improved modeling of the point spread function (PSF). Here, we summarize the cuts, which leave 146,756 events remaining in the data.  Although the original GRECO-Astronomy dataset is sensitive to all neutrino flavors, the cuts preferentially retain $\nu_\mu$. The first cut excludes events with angular uncertainties $>7^\circ$. Event direction reconstruction is performed for two nested hypotheses: a cascade hypothesis and a cascade plus track hypothesis. The second cut is on the likelihood ratio of these two hypotheses. Both of these cuts predominantly select tracks. Finally, we select events with at least one identifiable pulse ($>\,$0.25 photo-electrons) of unscattered light. This cut improves the biases of the likelihood used in section~\ref{sec:methods}. After cuts the median angular resolution is $\sim4^\circ$. See appendix~\ref{sec:grecocuts} for the effects of the cuts on angular resolution.

\subsection{X-Ray Bright AGN Selection} \label{subsec:agnselect}

We consider two source catalogs. Motivated by the intrinsic X-ray flux of NGC~1068, the first catalog is derived from the BAT AGN Spectroscopic Survey (BASS-DR-1) \citep{Ricci:2017ApJS}. We have defined a ranking of the AGN to isolate the most promising objects in the BASS catalog. Our methods described in section~\ref{sec:methods} cannot handle sources near the celestial poles, so we discard them outside the $\left[ -64.16^\circ, 70.05^\circ \right]$ declination range. Then, we calculate a proportionality factor between the energy flux of the 20-50 keV X-ray band and the differential neutrino particle flux of NGC~1068, and apply this factor to all objects in the BASS catalog in order to obtain an expected neutrino flux for each object. Assuming an $E^{-3}$ spectrum across all sources (including NGC 1068, as we believe the assumption is close enough to the measured value), we rank them via a metric defined to be the expected neutrino flux divided by the $3\sigma$ evidence potential and select the top 50 sources. The top-ranked source corresponds to NGC 4151, with NGC 1068 being the 16th ranked source. As classified in the BASS-DR1 catalog, the final set of AGN consists of 46 Seyfert-type galaxies, 2 FSRQs, and 2 BL-LACs.

\subsection{Galactic Source Selection} \label{subsec:galselect}

The second catalog consists of a selection of 56 Galactic sources observed by Fermi-LAT that have a spectral shape consistent with neutral pion decay \citep{Abdollahi2022Search}. We use the entire catalog as all objects lie within the declination range of our analysis. Additionally, we calculated the neutrino flux expectation for each source from the $\gamma$-ray spectra in the Fermi 4FGL-DR4 catalog \citep{4FGL}, assuming a maximally hadronic origin for the $\gamma$-rays and $p$-$p$ interactions following the calculations of \citealt{AHLERS201873}. The Cygnus cocoon \citep{Ackermann2011Cocoon}, 4FGL~J2028.6+4110e, has the largest signal expectation among the 56 objects with 2.6 expected neutrinos.

\section{Analysis Methods} \label{sec:methods}

This work uses an unbinned likelihood ratio hypothesis test based on reconstructed direction, energy and angular uncertainty of events \citep{Braun2008Methods}. Though many Galactic sources are extended, the point source approximation remains valid, as the source extension observed in $\gamma$-rays is usually smaller than the median angular uncertainty of events in our dataset.
We use a data-driven background estimation by randomizing (scrambling) the right ascension of events in the observed data, averaging out potential point sources and maintaining the statistical properties of the atmospheric muon and neutrino background. Time-integrated analysis using the GRECO-Astronomy dataset was not possible previously, as simulations and observed data disagreed on several reconstruction parameters, leading to problematic test statistic (TS) distributions. These issues disappear after applying the event selection described earlier.

In order to describe the spatial probability distribution function (PDF) of the signal, most IceCube analyses use the 3-dimensional von Mises-Fisher (3D-vMF) distribution \citep{Fisher1953Dispersion}:

\begin{equation}
\label{eq:psf}
\mathcal{S}_{\mathrm{space}} (\mathbf{x}_{i},\kappa_{i},\mathbf{x}_\mathrm{s}) = \frac{\kappa_i}{4\pi \sinh \kappa_i} \exp \left(\kappa_i \cos | \mathbf{x}_{i} - \mathbf{x}_\mathrm{s} |\right),
\end{equation}
where the parameter $\kappa_i$ quantifies the per-event localization. For small angular uncertainty, the 3D-vMF distribution approximates a 2D normal distribution with $\kappa_i \approx 1/\sigma_i^2$. However, the 3D-vMF distribution does not accurately describe DeepCore's PSF, which can lead to biases in the likelihood regression. 

\begin{figure*}[ht!]
\centering
\includegraphics[scale=0.75] {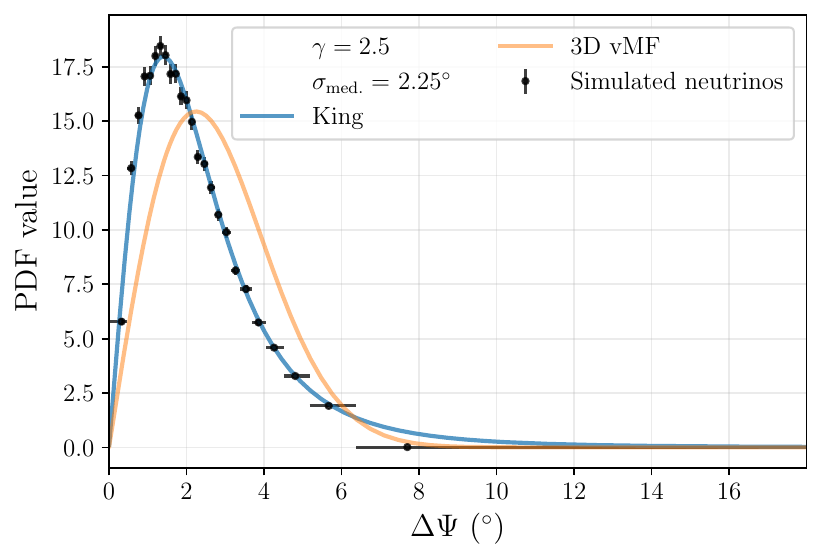}
\caption{A comparison between the King function and the 3D vMF expectation for simulated neutrinos with a median predicted angular error of 2.25 degrees. The simulation is shown in black, the King function fit in blue, and the 3D vMF in orange. The events in this ensemble are weighted to a spectral index of 2.5, with the predicted angular error of each event falling between 2.1 and 2.4 degrees.}
\label{fig:kingvsray}
\end{figure*}

Recent IceCube studies \citep{IceCube:2022Science, Icecube:2024Seyfert} have used Kernel Density Estimation (KDEs) \citep{KDE:Poluektov_2015} to describe the PSF. Although more accurate, KDEs are more computationally expensive. Instead, we adopt the King function \citep{1962AJ.....67..471K, Ackermann2013DETERMINATION} as an analytical and empirical formulation of the PSF. The King function is:

\begin{equation}
\label{eq:kings}
\mathcal{S}_{\mathrm{King}} (\mathbf{x}_{i},s_{i}, t_{i}, \mathbf{x}_\mathrm{s}) = \mathcal{N} \left( 1 + \frac{| \mathbf{x}_{i} - \mathbf{x}_\mathrm{s} |^2}{2 t_i s_i^2} \right)^{-t_i},
\end{equation}

\noindent with $\mathbf{x}_{i}$ representing the reconstructed direction of the $i$-th event, $\mathbf{x}_\mathrm{s}$ the direction of the point source being analyzed, $s_i$ and $t_i$ being per-event King function parameters, and $\mathcal{N}$ the overall normalization. Appendix~\ref{sec:kingnorm} describes novel computational methods for normalizing the King function. We determine $s_i$ and $t_i$ by fitting the King function to ensembles of simulated neutrino events, using penalized B-splines \citep{Eilers1996Flexible} to interpolate between fit points. This effectively makes the King function parameters a function of reconstruction variables, which can be determined on a per-event basis. This was done with respect to a single reconstruction variable, the reconstructed angular error $\sigma$, and for a fixed index of $\gamma = 2.5$, as we found this choice has minimal impact on $s_i$ and $t_i$. Figure~\ref{fig:kingvsray} shows an example comparison of the King function and the 3D-vMF parameterization.

The all-sky search is conducted between declinations $-64.16^\circ$ and $70.05^\circ$ segmenting the sky into pixels of equal area using HEALPix \citep{Gorski2005HEALPix}. The number of pixels in the sky is $N_\mathrm{pix}=12\times N_\mathrm{side}^2$. We adopt $N_\mathrm{side}=64$ corresponding to $N_\mathrm{pix}=49~152$ pixels. We only consider those pixels with centers in the declination range selected, reducing the number of pixels to 45~120. The coordinates of each pixel center are treated as a separate search location. For the second search, we use the two catalogs described in section~\ref{sec:data}. 

In a catalog search, we can identify the most likely neutrino emitter, but even the most significant source in each catalog may not be bright enough to produce a significant detection. To identify significant emission from a subset of catalog sources, we use the binomial test by minimizing \citep{Abbott2008Search}: 

\begin{equation}
\label{eq:binom}
P_k=\sum_{m=k}^N\frac{N!}{(N_m)!m!}p_k^m(1-p_k)^{N-m},
\end{equation}

\noindent with respect to $k$, where $p_k$ represents the local p-value for the $k$-th source obtained from the catalog search. To compute the binomial p-value for a catalog of $N$ sources, the local p-values of the sources are ordered, $p_1 \leq p_2 \leq \ldots \leq p_k \leq \ldots \leq p_N$. $P_k$ is then computed using eq.~\ref{eq:binom} for all $k\in[1,N]$, and the minimum value for $P_k$ represents the binomial p-value for the catalog with a sub-population of $k$ sources.

We take into account the look elsewhere effect, whereby looking at $N$ different locations leads the smallest p-value to be $\sim 1/N$ by random chance \citep{Bayer2020look}. We introduce a trials correction procedure and calculate a global p-value. For both the catalog search and the sky scan, we use the \v{S}id\'ak correction \citep{Sidak1967Rectangular} fit to the cumulative distribution of most significant p-values from a set of background scramblings. This is done instead of simply using the number sources in each case, as the relatively large angular uncertainty of the dataset combined with small inter-source spacing can give rise to correlations. For the binomial test, we use the cumulative distribution directly as it is poorly described by the Šidák correction.

\begin{figure*}[ht!]
\centering
\includegraphics[scale=0.75] {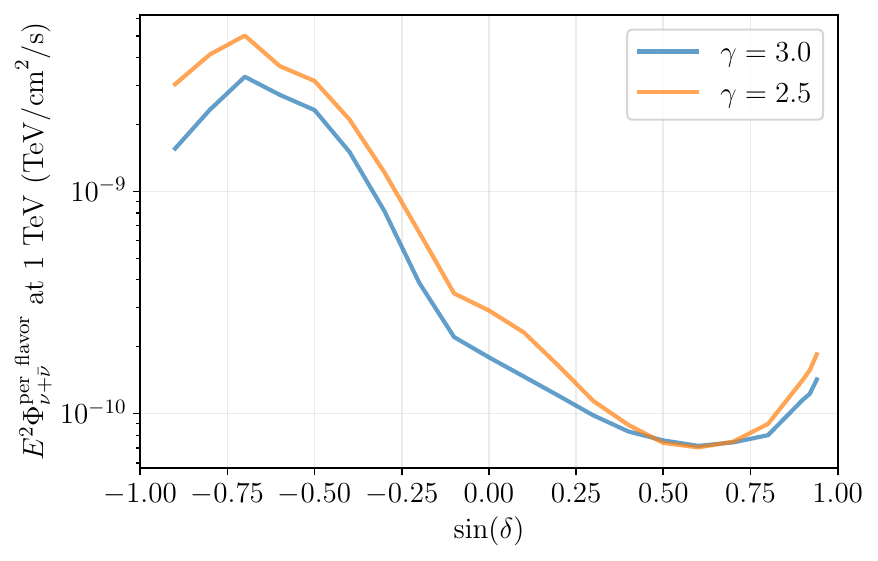}
\caption{Per-flavor $\nu + \bar{\nu}$ sensitivity for the time-integrated neutrino flux from a point source as a function of sine of declination for a power law spectrum of index 2.5 (orange) and 3.0 (blue). Systematic uncertainties have been considered following section~\ref{sec:systematics}.}
\label{fig:sens}
\end{figure*}

We benchmarked the performance of our methods by injecting simulated neutrinos from a source. We define \emph{sensitivity} as the flux required for which 90\% of the alternative hypothesis TS distribution exceeds the median of the null hypothesis TS distribution. 
The scrambling procedure results in a sensitivity that only depends on declination. The sensitivity of this analysis as a function of sine of declination for two spectral indices is shown in Figure~\ref{fig:sens}.

\section{Systematic Uncertainties} \label{sec:systematics}

Because we derive the background PDFs from data, we only need to consider the effect of systematic uncertainties in simulated neutrinos. The most important uncertainties are due to scattering and absorption of Cherenkov light in the South Pole ice, the relative DOM efficiency, and the properties of the re-frozen ice column around each string, known as \emph{hole ice}. We simulated additional datasets for $\pm 10\%$ variation in the scattering coefficient, $\pm 10\%$ variation in the absorption coefficient, $\pm 10\%$ variation in the relative DOM efficiency, and $\pm 1\sigma$ variation in hole-ice optical properties. The uncertainty in sensitivity due to these systematics for a power law index of $\gamma = 2.5$ and declination of $\delta = 0^\circ$ are: scattering 8\%, absorption 14\%, relative DOM efficiency 13\%, and hole ice 17\%. 
We add these uncertainties in quadrature for a total systematic uncertainty of 27\%. Thus, all limits and sensitivities presented here have been degraded by 27\%.

\section{Results} \label{sec:results}

\begin{figure*}[ht!]
\centering
\includegraphics[scale=0.35] {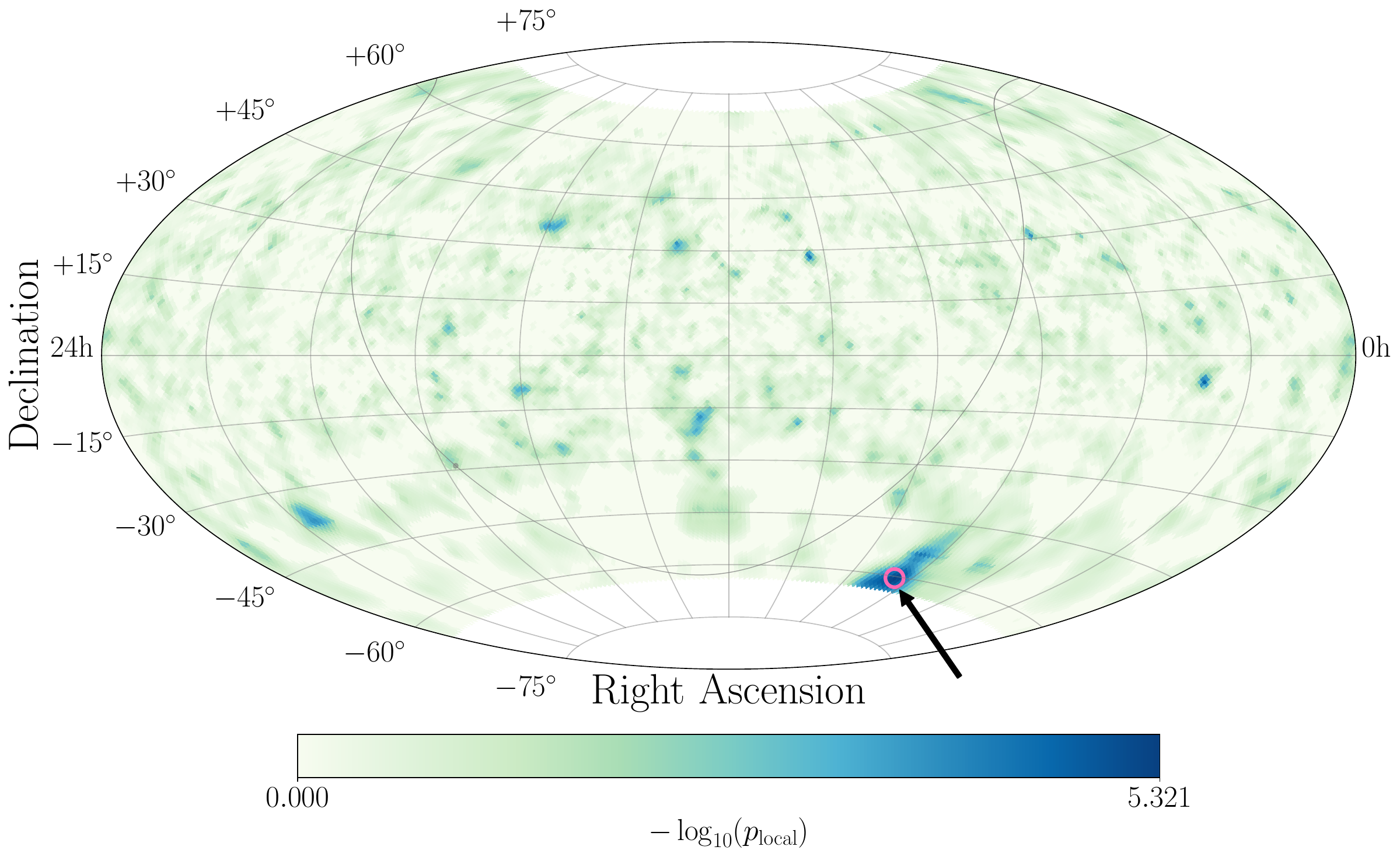}
\caption{Local p-value map in equatorial coordinates. The color scale represents the local p-value of each pixel in the sky, with the hottest spot highlighted by the pink circle and black arrow. Note the smearing effect at the southern declinations where the angular resolution is poor.}
\label{fig:skymap}
\end{figure*}

After performing the all-sky scan, we find the hottest spot at right ascension $\alpha = 97.86^\circ$ and declination $\delta = -60.43^\circ$, with a local p-value of $4.8 \times 10^{-6}$ (Figure~\ref{fig:skymap}). After accounting for the look-elsewhere effect, the global p-value is 0.078 ($1.4\sigma$). The coordinates of the hottest spot do not coincide with any object in either of the catalogs.

We find no evidence for neutrino emission from any of the 56 objects in the Galactic catalog. The most significant Galactic source is 4FGL J0240.5+6113, associated with LS I +61 303, a high-mass X-ray binary. The local p-value is $8 \times 10^{-3}$; and the global p-value is 0.30. No evidence for neutrino emission was found using the binomial test.
The most significant choice of $k$ yielded a local binomial p-value of 0.36, and a global p-value of 0.77. The best choice for $k$ was $k = 1$, with only 4FGL~J0240.5+6113 contributing. The results of the catalog and binomial search over Galactic sources are summarized in appendix~\ref{sec:appxbinom}.

Likewise, we find no evidence for neutrino emission from the 50 X-ray bright AGN in either the catalog search or the binomial test. Our catalog search identified Q0241+622 as the most significant AGN, with a local p-value of 0.049 and a global p-value of 0.89. NGC 1068 yielded the 23rd-most significant p-value, with a local p-value of 0.50, and a global p-value of 1. The binomial test preferred a sub-population of $k = 30$ sources with a local binomial p-value of 0.33 obtained from Eq.~\ref{eq:binom} and a global p-value of 0.83 after accounting for trials. The results of the catalog and binomial search over X-ray bright AGN are summarized in appendix~\ref{sec:appxbinom}. The most significant results from each search are summarized in Table~\ref{tab:results}.

\begin{table*}[ht!]
\centering
\caption{\label{tab:results}
Summary of the most significant results from both catalog searches and the all-sky scan. None yielded significant evidence for neutrino emission.
}
\begin{tabular}{lllll}
\toprule
Source catalog & Candidate name & $\alpha$ & $\delta$ & $p_\mathrm{local}$~($p_\mathrm{global}$) \\
\midrule
50 X-ray bright AGN & Q0241+622 & $41.24^\circ$ & $62.47^\circ$ & $4.9 \times 10^{-2}~(0.89)$ \\ 
56 Galactic sources & LS I +61 303 (4FGL~J0240.5+6113) & $40.14^\circ$ & $61.23^\circ$ & $8.0 \times 10^{-3}~(0.30)$ \\ 
All-sky scan & -- & $97.88^\circ$ & $-60.43^\circ$ & $4.8 \times 10^{-6}~(0.078)$ \\ 

\bottomrule

\end{tabular}

\end{table*}

\begin{figure*}[ht!]
\centering
\includegraphics[scale=.75] {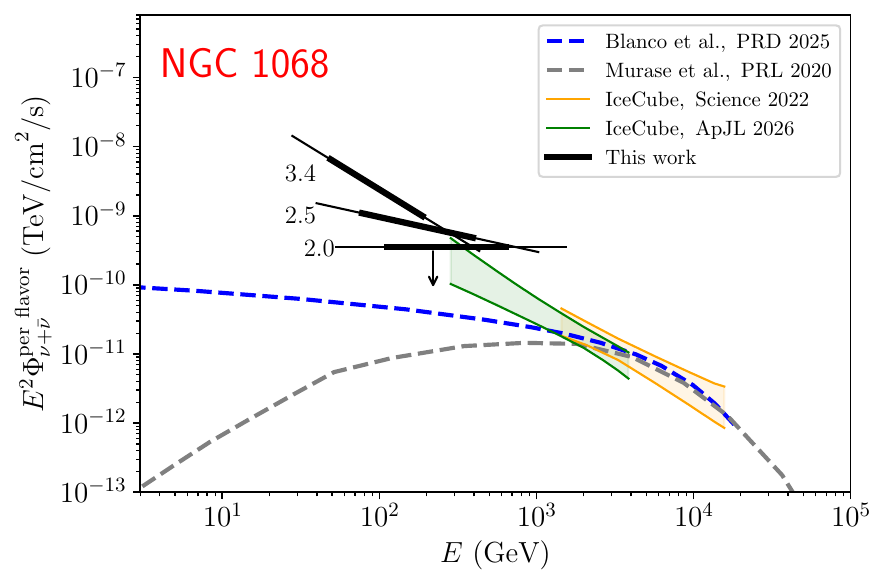}
\caption{Per-flavor $\nu+\bar{\nu}$ upper limits for NGC 1068 as a function of energy for three spectral indices (solid black). The thick black lines corresponds to the central 68\% energy range, while the thin black lines denote the central 95\% energy range. Models by \citealt{Blanco2025Neutrino} and \citealt{Murase:2020PhRvL} are shown in blue and gray, respectively. The best-fit flux by \citealt{IceCube:2022Science} and \citealt{abbasi2025evidenceneutrinoemissionxray} are shown in orange and green, respectively.}
\label{fig:NGC limit}
\end{figure*}

With no evidence for neutrino emission, we calculate 90\% confidence level (CL) upper limits on the time-integrated flux from the sources in both catalogs for power law spectral indices of $\gamma={2.5}$ and $\gamma={3.0}$. We also present the central 95\% energy range for each object.
Because there is evidence of higher-energy neutrino emission from NGC~1068, we include additional upper limits for NGC 1068 at spectral indices of $E^{-2.0}$ and $E^{-3.4}$, the latter being a constraint on the extrapolation of the high-energy spectrum. Upper limits for NGC 1068 are shown in Figure~\ref{fig:NGC limit} (see also appendix~\ref{sec:appxuladd}). The upper limit is a factor of $\sim 2$ higher than the extrapolation of the high-energy spectrum to lower energies. Limits for NGC~1068 assuming a power law with exponential cutoff are also provided in appendix~\ref{sec:appxuladd}. Additionally, upper limits on the \citealt{Murase:2020PhRvL} and \citealt{Blanco2025Neutrino} models are presented in appendix~\ref{sec:appxuladd}. Upper limits for all sources are summarized in appendix~\ref{sec:appxul}.

\section{Discussion} \label{sec:discuss}

We developed techniques to search for sub-TeV time-integrated neutrino emission from a catalog of Galactic sources and a catalog of intrinsically bright X-ray AGNs spanning almost the full sky. No evidence for neutrino emission is found for individual sources or for a subset of weak emitters. 

We limit the flux of NGC 1068 between 30 and 400 GeV to be at most a factor of 2 higher than the extrapolation of the high-energy \citep{abbasi2025evidenceneutrinoemissionxray} best fit, providing the strongest upper limit for time-integrated emission in this energy range. This compliments IceCube's high-energy focused analysis which detected neutrino emission from NGC 1068 as low as 0.3 TeV considering the 68\% energy interval.

Super-Kamiokande searched for $>$GeV neutrino point sources using all data prior to Gadolinium loading \citep{Wang2023Time}. Their sensitivity at the declination of NGC~1068 for a $\nu_\mu+\bar{\nu}_\mu$ flux integrated between 1~GeV and 10$^5$~GeV, assuming a spectral index of $\gamma=2.0$, is $\sim4.5\times 10^{-7}$ cm$^{-2}$~s$^{-1}$ at 90\% CL. The per-flavor limit presented here for the same index and energy integration range is $3.5\times 10^{-7}$ cm$^{-2}$~s$^{-1}$ at 90\% CL, but the sensitive energy range of Super-Kamiokande is lower than that of DeepCore. ANTARES has also searched for time-integrated neutrino emission from NGC~1068 \citep{2025arXiv251107239A} assuming a spectral index of $\gamma=2.5$. They constrain $\Phi_{\nu+\bar{\nu}}^{\mathrm{per\;flavor}} < 5.57\times10^{-11}$~TeV$^{-1}~$cm$^{-2}~$s$^{-1}$ at 90\% CL. This is comparable to the limits presented here, however, the 90\% central energy range for ANTARES at the declination of NGC 1068 is between 398 GeV and 316 TeV \citep{Albert2019ANTARESold}, which spans much higher energies compared to DeepCore.

Evidence or excess of neutrinos has been found for several $\gamma$-ray opaque galaxies. \citealt{abbasi2025evidenceneutrinoemissionxray} use a binomial test to find $3.3\sigma$ evidence for neutrino emission from a collection of 11 northern sky Seyferts, not including NGC~1068. \citealt{Abbasi2026EvidenceSouth} find $3\sigma$ evidence for a stacked search of neutrinos, with the strongest contributions from Circinus, NGC~7582, and ESO~138-1. \citealt{Neronov2024Neutrino} use IceCube public data to find $3\sigma$ evidence for neutrino emission from NGC~4151 and NGC~3079. \citealt{Abbasi2025SearchXR} find a 2.9$\sigma$ excess from NGC~4151. In most of these cases
the best fit spectrum is softer than an $E^{-2}$ power law; thus, limits presented here are astrophysically interesting. For the Cygnus cocoon, the limit is 7.6 times higher than the maximally hadronic neutrino flux derived from Fermi-LAT $\gamma$-ray observations (see Figure~\ref{fig:cyglimit}). 

New and upcoming detectors, such as the recently installed IceCube Upgrade \citep{Abbasi2026Physics} and KM3NeT ORCA \citep{Aiello2024Measurement} will improve sub-TeV sensitivity, with this analysis serving as a foundation for future time-integrated analyses conducted using the former. Improvements in event reconstruction algorithms using machine-learning methods \citep{Sgaard2023GraphNeT} also yield improvements in sensitivity. The case for sub-TeV measurements is particularly strong for NGC 1068, for which \citealt{abbasi2025evidenceneutrinoemissionxray} already provided evidence of sub-TeV emission. Future gamma-ray observations with instruments such as the Cherenkov Telescope Array Observatory \citep{Hofmann2022} will complement sub-TeV neutrino searches, further improving our understanding of sources with potentially soft neutrino spectra, including NGC 1068.

\begin{acknowledgments}
The authors gratefully acknowledge the support from the following agencies and institutions:
USA {\textendash} U.S. National Science Foundation-Office of Polar Programs,
U.S. National Science Foundation-Physics Division,
U.S. National Science Foundation-EPSCoR,
U.S. National Science Foundation-Office of Advanced Cyberinfrastructure,
Wisconsin Alumni Research Foundation,
Center for High Throughput Computing (CHTC) at the University of Wisconsin{\textendash}Madison,
Open Science Grid (OSG),
Partnership to Advance Throughput Computing (PATh),
Advanced Cyberinfrastructure Coordination Ecosystem: Services {\&} Support (ACCESS),
Frontera and Ranch computing project at the Texas Advanced Computing Center,
U.S. Department of Energy-National Energy Research Scientific Computing Center,
Particle astrophysics research computing center at the University of Maryland,
Michigan State University,
Astroparticle physics computational facility at Marquette University,
NVIDIA Corporation,
and Google Cloud Platform;
Belgium {\textendash} Funds for Scientific Research (FRS-FNRS and FWO),
FWO Odysseus and Big Science programmes,
and Belgian Federal Science Policy Office (Belspo);
Germany {\textendash} Bundesministerium f{\"u}r Forschung, Technologie und Raumfahrt (BMFTR),
Deutsche Forschungsgemeinschaft (DFG),
Helmholtz Alliance for Astroparticle Physics (HAP),
Initiative and Networking Fund of the Helmholtz Association,
Deutsches Elektronen Synchrotron (DESY),
and High Performance Computing cluster of the RWTH Aachen;
Sweden {\textendash} Swedish Research Council,
Swedish Polar Research Secretariat,
Swedish National Infrastructure for Computing (SNIC),
and Knut and Alice Wallenberg Foundation;
European Union {\textendash} EGI Advanced Computing for research;
Australia {\textendash} Australian Research Council;
Canada {\textendash} Natural Sciences and Engineering Research Council of Canada,
Calcul Qu{\'e}bec, Compute Ontario, Canada Foundation for Innovation, WestGrid, and Digital Research Alliance of Canada;
Denmark {\textendash} Villum Fonden, Carlsberg Foundation, and European Commission;
New Zealand {\textendash} Marsden Fund;
Japan {\textendash} Japan Society for Promotion of Science (JSPS), Ministry of Education, Culture, Sports, Science and Technology (MEXT), and Institute for Global Prominent Research (IGPR) of Chiba University;
Korea {\textendash} National Research Foundation of Korea (NRF);
Switzerland {\textendash} Swiss National Science Foundation (SNSF).
\end{acknowledgments}

%
\facility{IceCube}




\appendix

\section{Effect of Cuts on Angular Resolution} \label{sec:grecocuts}

\begin{figure*}[ht!]
\centering
\includegraphics[scale=.75] {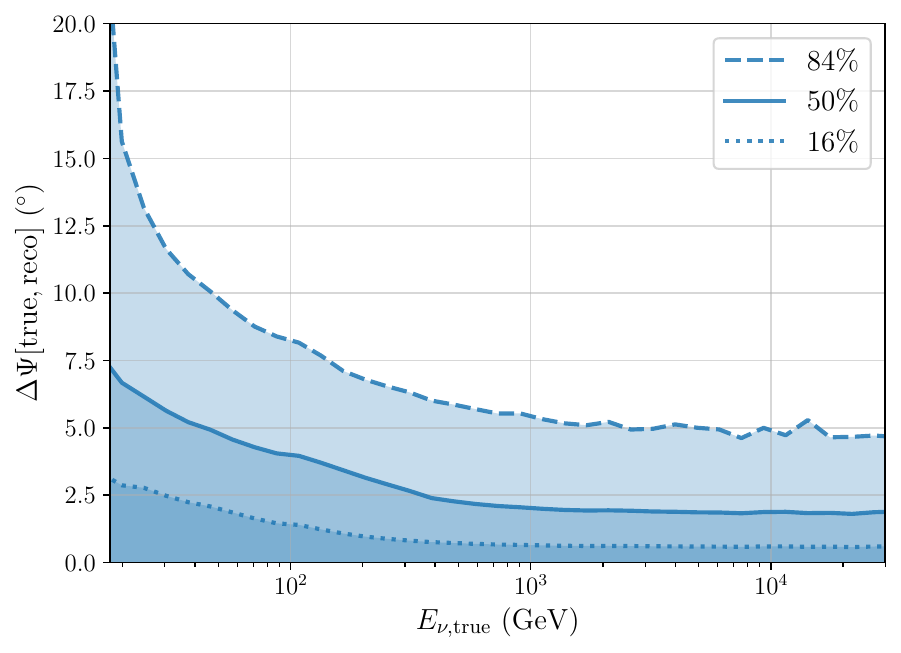}
\caption{Distribution of the angular separation between reconstructed event direction and true neutrino direction, as a function of true neutrino energy. Three confidence levels are shown (16\%, 50\%, and 84\%), with all events being weighted to an $E^{-2}$ power law spectrum. Compare with Figure 10 in \citealt{Abbasi2023Search}.}
\label{fig:angerrcut}
\end{figure*}

This study uses a modified version of the GRECO-Astronomy (GeV Reconstructed Events with Containment for Oscillations) event selection, first developed for the tau neutrino appearance analysis in \citealt{Aartsen2019MeasurementNuTau}. In order to enable time-integrated analysis, we apply direct cuts to the dataset, as described in subsection~\ref{subsec:detdata}. This has the effect of improving the overall angular resolution of the dataset, as shown in Figure~\ref{fig:angerrcut}, compared to the uncut dataset.

\section{Normalization of the King Function} \label{sec:kingnorm}

Any valid expression for the signal spatial PDF should respect the following normalization condition:

\begin{equation}\label{eq:spacenorm}
    \int_S \mathcal{S}_\mathrm{Space} d \Omega = 1,
\end{equation}
where $S$ is the unit sphere and $d\Omega$ is the differential solid angle element defined as $d\Omega = \sin \Psi d\Psi d\phi$. Thus, as it appears in eq.~\ref{eq:kings}, the King function must satisfy:

\begin{equation}\label{eq:kingreq}
    \int_0^{2\pi} \int_0^{\pi} \mathcal{N} \left( 1 + \frac{\Psi^2}{2 t s^2} \right)^{-t} \sin \Psi d\Psi d\phi = 1,
\end{equation}

Solving for the normalization constant $\mathcal{N}$ yields:
\begin{equation}\label{eq:king_norm}
    \mathcal{N} = \frac{1}{2\pi \eta}; \quad \text{with~} \eta(s,t) = \int_0^{\pi} \left( 1 + \frac{\Psi^2}{2 t s^2} \right)^{-t} \sin \Psi d\Psi,
\end{equation}
which admits an exact solution as a rapidly convergent alternating series. 
Substituting the Taylor series for $\sin(\Psi)$ around $\Psi=0$, we integrate term-by-term since the series is uniformly convergent \citep{rudin_1976_principles}. This yields:
\begin{equation}\label{eq:king_series}
   \eta(s,t) = \sum_{n=0}^{\infty} (-1)^n a_n; \quad \text{with~} a_n =\frac{ \pi^{2n+2}}{2(n+1)(2n+1)!} 
   {}_2F_1\!\left(n{+}1, t; n{+}2; {-}\frac{\pi^2}{2ts^2}\right),
\end{equation}
where ${}_2F_1(a,b;c;z)$ is the Gauss hypergeometric function.




For the sake of computational speed, this analysis uses two approximations for eq.~\ref{eq:king_norm}. For a combination of small $s$ and large $t$, the integrand of eq.~\ref{eq:king_norm} is sharply peaked near zero, allowing us to invoke a small-angle approximation $d\Omega \approx \Psi d \Psi d \phi$. This produces a simple analytical solution for the normalization:

\begin{equation}\label{eq:kingsma}
  \mathcal{N} \approx \frac{1}{2 \pi s^2} \left( 1 - \frac{1}{t} \right),
\end{equation}

Under the conditions where $\Psi \gg 0.1^{\circ}$, we evaluate eq.~\ref{eq:king_norm} using a Gauss-Legendre quadrature, along with the change of variable of $\Psi \to \cos \Psi$. This gives an approximation of the form:

\begin{equation}\label{eq:kingsglq}
  \eta(s,t) = \int_{-1}^1 \left( 1 + \frac{\arccos^2\Psi}{2 t s^2} \right)^{-t} d\Psi \approx \sum_{i = 1}^n w_i \left( 1 + \frac{\arccos^2\Psi_i}{2 t s^2} \right)^{-t},
\end{equation}

\noindent where $n$ is the number of sample points used, $\Psi_i$ is the $i$-th root of the $n$-th Legendre polynomial $P_n(\Psi)$, and $w_i$ is the $i$-th sample point defined as:

\begin{equation}\label{eq:glqweight}
  w_i = \frac{2}{\left( 1 - \Psi_i^2 \right) \left[P'_n(\Psi_i)\right]^2},
\end{equation}

The usage of either approximation is determined by comparing event-by-event $s$ and $t$ values against set thresholds for $s$ and $t$. This analysis sets the thresholds for $s=2.5^\circ$ and $t = 2$.

\section{Binomial Search Results} \label{sec:appxbinom}

The following tables present the full results of the binomial tests performed on both X-ray AGN and pion-bump source catalogs. Table~\ref{tab:binomresults} summarizes the results of the binomial test for each catalog, while Table~\ref{tab:binomresultsdetails} details the attributes for each of the $k$ sources corresponding to each catalog's optimized binomial p-value. The sources in each table are listed in order of ascending $p_\mathrm{local}$ value.

\begin{table*}[ht!]
\centering
\caption{\label{tab:binomresults}
Summary of binomial test results for both Galactic and extragalactic source catalogs. For each catalog, we list the value of $k$ that corresponds to the optimized binomial p-value, along with the trial-corrected global p-value.
}
\begin{tabular}{lll}
\toprule
Source catalog & $k$ & $P_{k,\mathrm{local}}$~($P_{k,\mathrm{global}}$) \\
\midrule
50 X-ray bright AGN & 30 & 0.33 (0.83) \\
56 Galactic sources & 1 & 0.36 (0.77) \\ 
\bottomrule
\end{tabular}
\end{table*}

\pagebreak

\begin{longtable}{lllll}
\caption{\label{tab:binomresultsdetails}
Properties of the $k$ top sources selected by the binomial test for each catalog. We list the name, right ascension, declination, local p-value, and the trial-corrected global p-value for each source.
}
\\
\toprule
Source catalog & Candidate name & $\alpha$ & $\delta$ & $p_\mathrm{local}$~($p_\mathrm{global}$) \\
\midrule
50 X-ray bright AGN & -- & -- & -- & -- \\
~ & Q0241+622 & $41.24^\circ$ & $62.47^\circ$ & 0.049 (0.89) \\
~ & NGC 7319 & $339.01^\circ$ & $33.98^\circ$ & 0.093 (--) \\
~ & 4C +50.55 & $321.16^\circ$ & $50.97^\circ$ & 0.11 (--) \\
~ & Mrk 348 & $12.20^\circ$ & $31.96^\circ$ & 0.11 (--) \\
~ & NGC 4945 & $196.36^\circ$ & $-49.47^\circ$ & 0.12 (--) \\
~ & LEDA 166445 & $42.68^\circ$ & $54.70^\circ$ & 0.14 (--) \\
~ & Mrk 1498 & $247.02^\circ$ & $51.78^\circ$ & 0.15 (--) \\
~ & NGC 1275 & $49.95^\circ$ & $41.51^\circ$ & 0.15 (--) \\
~ & IRAS 05078+1626 & $77.69^\circ$ & $16.50^\circ$ & 0.16 (--) \\
~ & Mrk 1040 & $37.06^\circ$ & $31.31^\circ$ & 0.19 (--) \\
~ & UGC 3374 & $88.72^\circ$ & $46.44^\circ$ & 0.19 (--) \\
~ & NGC 7469 & $345.82^\circ$ & $8.87^\circ$ & 0.22 (--) \\
~ & NGC 4992 & $197.27^\circ$ & $11.63^\circ$ & 0.25 (--) \\
~ & LEDA 138501 & $32.41^\circ$ & $52.44^\circ$ & 0.25 (--) \\
~ & NGC 1194 & $45.95^\circ$ & $-1.10^\circ$ & 0.27 (--) \\
~ & NGC 4151 & $182.64^\circ$ & $39.41^\circ$ & 0.29 (--) \\
~ & MCG +4-48-2 & $307.15^\circ$ & $25.73^\circ$ & 0.31 (--) \\
~ & Mrk 417 & $162.38^\circ$ & $22.96^\circ$ & 0.33 (--) \\
~ & Z164-19 & $221.40^\circ$ & $27.03^\circ$ & 0.39 (--) \\
~ & NGC 5506 & $213.31^\circ$ & $-3.21^\circ$ & 0.47 (--) \\
~ & IRAS 05589+2828 & $90.54^\circ$ & $28.47^\circ$ & 0.49 (--) \\
~ & 3C 273 & $187.28^\circ$ & $2.05^\circ$ & 0.49 (--) \\
~ & NGC 1068 & $40.67^\circ$ & $-0.01^\circ$ & 0.50 (--) \\
~ & Mrk 501 & $253.47^\circ$ & $39.76^\circ$ & 0.51 (--) \\
~ & NGC 2110 & $88.05^\circ$ & $-7.46^\circ$ & 0.52 (--) \\
~ & Mrk 421 & $166.11^\circ$ & $38.21^\circ$ & 0.52 (--) \\
~ & NGC 4102 & $181.60^\circ$ & $52.71^\circ$ & 0.52 (--) \\
~ & NGC 1142 & $43.80^\circ$ & $-0.18^\circ$ & 0.52 (--) \\
~ & 3C 111 & $64.59^\circ$ & $38.03^\circ$ & 0.56 (--) \\
~ & NGC 3079 & $150.49^\circ$ & $55.68^\circ$ & 0.56 (--) \\
\midrule

56 Galactic sources & -- & -- & -- & -- \\
~ & LS I +61 303 (4FGL~J0240.5+6113) & $40.14^\circ$ & $61.23^\circ$ & $8.0 \times 10^{-3}~(0.30)$ \\

\bottomrule
\end{longtable}


\section{Additional Upper Limits} \label{sec:appxuladd}

\begin{figure*}[ht!]
\centering
\includegraphics[scale=.75] {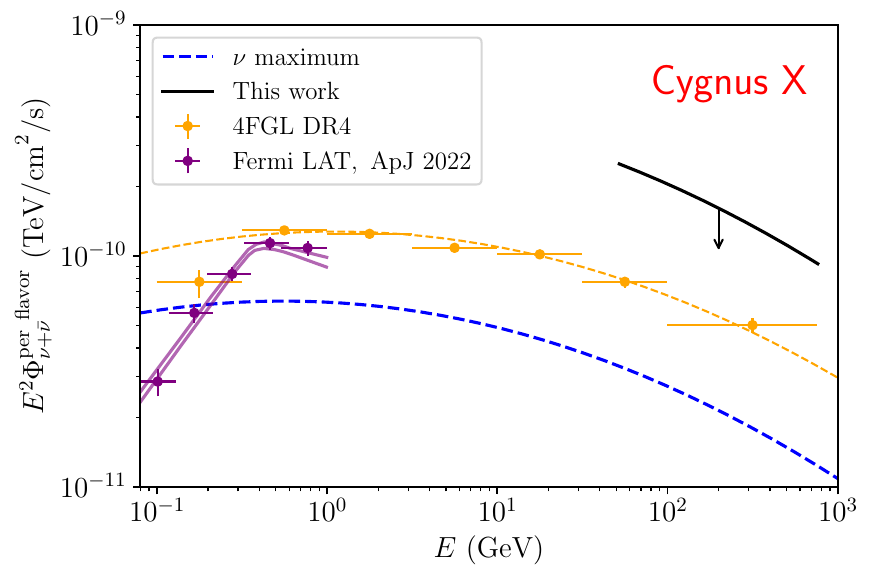}
\caption{Per-flavor $\nu+\bar{\nu}$ upper limits for the Cygnus cocoon as a function of energy for a maximally hadronic log-parabola spectrum extrapolated from \citealt{4FGL} following \citealt{AHLERS201873} in black. The dotted yellow line and error bars represent the $\gamma$-ray spectrum as reported in \citealt{4FGL}. The dashed blue line represents the maximum possible neutrino spectrum extrapolated from the $\gamma$-ray spectrum. Finally, the purple error band represents the results from the pion-bump analysis as reported in \citealt{Abdollahi2022Search}. The limit presented here is drawn within the central 95\% energy range.}
\label{fig:cyglimit}
\end{figure*}

We have included additional upper limits for NGC 1068 to aid in the development of theoretical models. We present power law with exponential cutoff upper limits in Table~\ref{tab:NGC plco} for a variety of spectral indices and cutoff energies. In Table~\ref{tab:NGC addpl} we provide the numerical values for the limits presented in Figure~\ref{fig:NGC limit} for values of $\gamma$ not already present in Table~\ref{tab:catabassresults}. We also provide upper limits on the models described in \citealt{Murase:2020PhRvL} and \citealt{Blanco2025Neutrino} in Table~\ref{tab:NGC modadd}. Finally, we provide an additional upper limit for the Cygnus cocoon in Figure~\ref{fig:cyglimit}, given its status as a known high-energy $\gamma$-ray source and likely cosmic-ray source.

\begin{longtable}{llllll}
\caption{\label{tab:NGC plco}
Additional upper limits for NGC 1068 assuming a source spectrum following a power law with exponential cutoff of the form $\Phi(E) = \Phi_0 (E / 1~\mathrm{TeV})^{-\gamma} \exp{\left( -E / E_\mathrm{cutoff} \right)}$. We list the spectral index, the cutoff energy in GeV, the 90\% CL upper limits on the number of source neutrinos as well as the astrophysical $\nu + \bar{\nu}$ per-flavor flux evaluated at 1 TeV with units of TeV$^{-1}$ cm$^{-2}$ s$^{-1}$, and the energies that bracket the 95\% central energy range for each respective limit in GeV.
}
\\
\toprule
$\gamma$ & $E_\mathrm{cutoff}$ & $n_{\mathrm{s},90\%}$ & $\Phi |_{1\mathrm{TeV}}$ & $E_{2.5\%}$ & $E_{97.5\%}$ \\
\midrule
2.0 & 100  & 39.3 & $4.2 \times 10^{-13}$ & 32  & 272  \\
2.0 & 177  & 33.4 & $1.5 \times 10^{-11}$ & 35  & 367  \\
2.0 & 316  & 29.0 & $9.7 \times 10^{-11}$ & 39  & 494  \\
2.0 & 562  & 24.7 & $2.3 \times 10^{-10}$ & 42  & 649  \\
2.0 & 1000  & 21.7 & $3.3 \times 10^{-10}$ & 45  & 853  \\
2.0 & 1778  & 19.5 & $3.8 \times 10^{-10}$ & 47  & 1016  \\
2.0 & 3162  & 18.0 & $3.9 \times 10^{-10}$ & 49  & 1151  \\
2.0 & 5623  & 17.2 & $3.8 \times 10^{-10}$ & 51  & 1226  \\
2.0 & 10000  & 16.9 & $3.8 \times 10^{-10}$ & 51  & 1327  \\
2.1 & 100  & 40.0 & $3.4 \times 10^{-13}$ & 31  & 264  \\
2.1 & 177  & 34.4 & $1.3 \times 10^{-11}$ & 34  & 354  \\
2.1 & 316  & 30.0 & $8.4 \times 10^{-11}$ & 37  & 474  \\
2.1 & 562  & 25.6 & $2.0 \times 10^{-10}$ & 40  & 618  \\
2.1 & 1000  & 23.1 & $3.1 \times 10^{-10}$ & 43  & 795  \\
2.1 & 1778  & 20.7 & $3.6 \times 10^{-10}$ & 45  & 952  \\
2.1 & 3162  & 19.2 & $3.7 \times 10^{-10}$ & 46  & 1101  \\
2.1 & 5623  & 18.3 & $3.7 \times 10^{-10}$ & 47  & 1168  \\
2.1 & 10000  & 17.5 & $3.6 \times 10^{-10}$ & 48  & 1229  \\
2.2 & 100  & 40.1 & $2.7 \times 10^{-13}$ & 31  & 256  \\
2.2 & 177  & 35.2 & $1.1 \times 10^{-11}$ & 33  & 341  \\
2.2 & 316  & 30.9 & $7.2 \times 10^{-11}$ & 36  & 450  \\
2.2 & 562  & 26.9 & $1.8 \times 10^{-10}$ & 39  & 599  \\
2.2 & 1000  & 24.2 & $2.8 \times 10^{-10}$ & 41  & 744  \\
2.2 & 1778  & 21.8 & $3.3 \times 10^{-10}$ & 43  & 907  \\
2.2 & 3162  & 20.1 & $3.5 \times 10^{-10}$ & 44  & 1019  \\
2.2 & 5623  & 19.3 & $3.5 \times 10^{-10}$ & 45  & 1126  \\
2.2 & 10000  & 19.0 & $3.6 \times 10^{-10}$ & 45  & 1166  \\
2.3 & 100  & 42.3 & $2.2 \times 10^{-13}$ & 29  & 250  \\
2.3 & 177  & 36.5 & $9.0 \times 10^{-12}$ & 32  & 330  \\
2.3 & 316  & 32.0 & $6.1 \times 10^{-11}$ & 35  & 431  \\
2.3 & 562  & 28.2 & $1.6 \times 10^{-10}$ & 37  & 568  \\
2.3 & 1000  & 25.3 & $2.5 \times 10^{-10}$ & 39  & 695  \\
2.3 & 1778  & 23.3 & $3.1 \times 10^{-10}$ & 41  & 852  \\
2.3 & 3162  & 21.9 & $3.3 \times 10^{-10}$ & 42  & 951  \\
2.3 & 5623  & 20.7 & $3.4 \times 10^{-10}$ & 42  & 1037  \\
2.3 & 10000  & 20.1 & $3.4 \times 10^{-10}$ & 43  & 1108  \\
2.4 & 100  & 41.7 & $1.7 \times 10^{-13}$ & 29  & 243  \\
2.4 & 177  & 37.5 & $7.4 \times 10^{-12}$ & 32  & 318  \\
2.4 & 316  & 33.6 & $5.3 \times 10^{-11}$ & 34  & 415  \\
2.4 & 562  & 29.1 & $1.4 \times 10^{-10}$ & 36  & 531  \\
2.4 & 1000  & 25.9 & $2.2 \times 10^{-10}$ & 37  & 651  \\
2.4 & 1778  & 24.5 & $2.8 \times 10^{-10}$ & 39  & 778  \\
2.4 & 3162  & 22.7 & $3.0 \times 10^{-10}$ & 39  & 892  \\
2.4 & 5623  & 22.4 & $3.2 \times 10^{-10}$ & 40  & 959  \\
2.4 & 10000  & 21.4 & $3.2 \times 10^{-10}$ & 41  & 1024  \\
2.5 & 100  & 42.1 & $1.4 \times 10^{-13}$ & 28  & 236  \\
2.5 & 177  & 38.1 & $6.0 \times 10^{-12}$ & 31  & 305  \\
2.5 & 316  & 34.2 & $4.4 \times 10^{-11}$ & 33  & 398  \\
2.5 & 562  & 30.9 & $1.2 \times 10^{-10}$ & 35  & 501  \\
2.5 & 1000  & 27.8 & $2.0 \times 10^{-10}$ & 36  & 617  \\
2.5 & 1778  & 25.6 & $2.5 \times 10^{-10}$ & 37  & 723  \\
2.5 & 3162  & 24.5 & $2.8 \times 10^{-10}$ & 38  & 837  \\
2.5 & 5623  & 23.5 & $2.9 \times 10^{-10}$ & 38  & 900  \\
2.5 & 10000  & 23.0 & $3.0 \times 10^{-10}$ & 38  & 951  \\
\bottomrule
\end{longtable}

\begin{table*}[ht!]
\centering
\caption{\label{tab:NGC addpl}
Additional power law upper limits for NGC 1068. We list the spectral index, the 90\% CL upper limits on the number of source neutrinos as well as the astrophysical $\nu + \bar{\nu}$ per-flavor flux evaluated at 1 TeV with units of TeV$^{-1}$ cm$^{-2}$ s$^{-1}$, and the energies that bracket the 95\% central energy range for each respective limit in GeV.
}
\begin{tabular}{lllll}
\toprule
$\gamma$ & $n_{\mathrm{s},90\%}$ & $\Phi |_{1\mathrm{TeV}}$ & $E_{2.5\%}$ & $E_{97.5\%}$ \\
\midrule
2.0 & 16.0 & $3.5 \times 10^{-10}$ & 53  & 1546  \\
3.2 & 33.4 & $1.3 \times 10^{-10}$ & 29  & 524  \\
3.4 & 36.9 & $9.5 \times 10^{-11}$ & 27  & 432  \\
\bottomrule
\end{tabular}
\end{table*}

\begin{table*}[ht!]
\centering
\caption{\label{tab:NGC modadd}
Additional upper limits for NGC 1068 assuming emission models described in \citealt{Murase:2020PhRvL} and \citealt{Blanco2025Neutrino}, respectively.  We list the 90\% CL upper limits on the number of source neutrinos as well as the model normalization, and the energies that bracket the 95\% central energy range for each respective limit in GeV.
}
\begin{tabular}{lllll}
\toprule
Reference & $n_{\mathrm{s},90\%}$ & Model normalization & $E_{2.5\%}$ & $E_{97.5\%}$ \\
\midrule
\citealt{Murase:2020PhRvL} & 15.1 & 29.1 & 65  & 1370  \\
\citealt{Blanco2025Neutrino} & 19.9 & 14.6 & 44  & 1103  \\
\bottomrule
\end{tabular}
\end{table*}

\section{Catalog Search Upper Limits} \label{sec:appxul}

The following tables summarize the upper limits placed on each source for each catalog search. The sources in each catalog are ordered by decreasing local significance. Table~\ref{tab:catabassresults} presents the results of the X-ray AGN catalog, while Table~\ref{tab:catapbresults} presents the results of the pion-bump catalog.

\begin{longtable}{lllllllll}
\caption{\label{tab:catabassresults}
Summary of the catalog search done on a selection of 50 X-ray bright AGN. Sources are ordered in descending local significance. For each source, we list the equatorial coordinates (J2000 equinox) as listed in the BASS DR1 catalog, the local (global) p-value, the spectral index for the corresponding upper limit, the 90\% CL upper limits on the number of source neutrinos as well as the astrophysical $\nu + \bar{\nu}$ per-flavor flux evaluated at 1 TeV with units of TeV$^{-1}$ cm$^{-2}$ s$^{-1}$, and the energies that bracket the 95\% central energy range for each respective limit in GeV.
}
\\
\toprule
Candidate name & $\alpha$ & $\delta$ & $p_\mathrm{local}$~($p_\mathrm{global}$) & $\gamma$ & $n_{\mathrm{s},90\%}$ & $\Phi |_{1\mathrm{TeV}}$ & $E_{2.5\%}$ & $E_{97.5\%}$ \\
\midrule
Q0241+622 & $41.24^\circ$ & $62.47^\circ$ & 0.05 (0.89) & 2.5 & 50.9 & $3.4 \times 10^{-10}$ & 50  & 646  \\
-- & -- & -- & -- & 3.0 & 69.4 & $2.7 \times 10^{-10}$ & 39  & 469  \\
NGC 7319 & $339.01^\circ$ & $33.98^\circ$ & 0.09 (--) & 2.5 & 31.4 & $1.5 \times 10^{-10}$ & 46  & 772  \\
-- & -- & -- & -- & 3.0 & 48.5 & $1.5 \times 10^{-10}$ & 33  & 550  \\
4C +50.55 & $321.16^\circ$ & $50.97^\circ$ & 0.11 (--) & 2.5 & 32.2 & $1.7 \times 10^{-10}$ & 50  & 678  \\
-- & -- & -- & -- & 3.0 & 46.2 & $1.5 \times 10^{-10}$ & 36  & 470  \\
Mrk 348 & $12.20^\circ$ & $31.96^\circ$ & 0.11 (--) & 2.5 & 29.0 & $1.4 \times 10^{-10}$ & 45  & 746  \\
-- & -- & -- & -- & 3.0 & 45.5 & $1.4 \times 10^{-10}$ & 33  & 537  \\
NGC 4945 & $196.36^\circ$ & $-49.47^\circ$ & 0.12 (--) & 2.5 & 22.2 & $6.7 \times 10^{-9}$ & 32  & 952  \\
-- & -- & -- & -- & 3.0 & 36.3 & $4.4 \times 10^{-9}$ & 25  & 585  \\
LEDA 166445 & $42.68^\circ$ & $54.70^\circ$ & 0.14 (--) & 2.5 & 30.9 & $1.8 \times 10^{-10}$ & 50  & 665  \\
-- & -- & -- & -- & 3.0 & 44.5 & $1.6 \times 10^{-10}$ & 37  & 465  \\
Mrk 1498 & $247.02^\circ$ & $51.78^\circ$ & 0.15 (--) & 2.5 & 31.0 & $1.6 \times 10^{-10}$ & 50  & 676  \\
-- & -- & -- & -- & 3.0 & 44.2 & $1.5 \times 10^{-10}$ & 36  & 469  \\
NGC 1275 & $49.95^\circ$ & $41.51^\circ$ & 0.15 (--) & 2.5 & 29.3 & $1.4 \times 10^{-10}$ & 48  & 723  \\
-- & -- & -- & -- & 3.0 & 42.6 & $1.3 \times 10^{-10}$ & 35  & 515  \\
IRAS 05078+1626 & $77.69^\circ$ & $16.50^\circ$ & 0.16 (--) & 2.5 & 33.8 & $2.2 \times 10^{-10}$ & 44  & 893  \\
-- & -- & -- & -- & 3.0 & 48.0 & $1.8 \times 10^{-10}$ & 33  & 567  \\
Mrk 1040 & $37.06^\circ$ & $31.31^\circ$ & 0.19 (--) & 2.5 & 25.3 & $1.3 \times 10^{-10}$ & 45  & 735  \\
-- & -- & -- & -- & 3.0 & 40.1 & $1.3 \times 10^{-10}$ & 33  & 535  \\
UGC 3374 & $88.72^\circ$ & $46.44^\circ$ & 0.19 (--) & 2.5 & 27.1 & $1.3 \times 10^{-10}$ & 49  & 698  \\
-- & -- & -- & -- & 3.0 & 39.6 & $1.3 \times 10^{-10}$ & 35  & 506  \\
NGC 7469 & $345.82^\circ$ & $8.87^\circ$ & 0.22 (--) & 2.5 & 34.7 & $3.2 \times 10^{-10}$ & 41  & 990  \\
-- & -- & -- & -- & 3.0 & 47.2 & $2.2 \times 10^{-10}$ & 32  & 629  \\
NGC 4992 & $197.27^\circ$ & $11.63^\circ$ & 0.25 (--) & 2.5 & 32.6 & $2.5 \times 10^{-10}$ & 43  & 873  \\
-- & -- & -- & -- & 3.0 & 44.6 & $1.9 \times 10^{-10}$ & 31  & 566  \\
LEDA 138501 & $32.41^\circ$ & $52.44^\circ$ & 0.25 (--) & 2.5 & 26.3 & $1.4 \times 10^{-10}$ & 50  & 688  \\
-- & -- & -- & -- & 3.0 & 36.9 & $1.2 \times 10^{-10}$ & 36  & 475  \\
NGC 1194 & $45.95^\circ$ & $-1.10^\circ$ & 0.27 (--) & 2.5 & 31.6 & $4.5 \times 10^{-10}$ & 38  & 1095  \\
-- & -- & -- & -- & 3.0 & 43.9 & $2.7 \times 10^{-10}$ & 32  & 638  \\
NGC 4151 & $182.64^\circ$ & $39.41^\circ$ & 0.29 (--) & 2.5 & 22.4 & $1.1 \times 10^{-10}$ & 47  & 721  \\
-- & -- & -- & -- & 3.0 & 32.8 & $1.0 \times 10^{-10}$ & 34  & 509  \\
MCG +4-48-2 & $307.15^\circ$ & $25.73^\circ$ & 0.31 (--) & 2.5 & 22.0 & $1.2 \times 10^{-10}$ & 45  & 790  \\
-- & -- & -- & -- & 3.0 & 33.6 & $1.1 \times 10^{-10}$ & 31  & 550  \\
Mrk 417 & $162.38^\circ$ & $22.96^\circ$ & 0.33 (--) & 2.5 & 20.7 & $1.2 \times 10^{-10}$ & 45  & 815  \\
-- & -- & -- & -- & 3.0 & 33.5 & $1.1 \times 10^{-10}$ & 32  & 556  \\
Z164-19 & $221.40^\circ$ & $27.03^\circ$ & 0.39 (--) & 2.5 & 18.8 & $9.9 \times 10^{-11}$ & 45  & 789  \\
-- & -- & -- & -- & 3.0 & 29.6 & $9.8 \times 10^{-11}$ & 32  & 545  \\
NGC 5506 & $213.31^\circ$ & $-3.21^\circ$ & 0.47 (--) & 2.5 & 21.4 & $3.3 \times 10^{-10}$ & 38  & 1086  \\
-- & -- & -- & -- & 3.0 & 31.1 & $2.1 \times 10^{-10}$ & 31  & 650  \\
IRAS 05589+2828 & $90.54^\circ$ & $28.47^\circ$ & 0.49 (--) & 2.5 & 15.5 & $8.0 \times 10^{-11}$ & 45  & 865  \\
-- & -- & -- & -- & 3.0 & 24.3 & $8.0 \times 10^{-11}$ & 32  & 545  \\
3C 273 & $187.28^\circ$ & $2.05^\circ$ & 0.49 (--) & 2.5 & 22.6 & $2.8 \times 10^{-10}$ & 38  & 991  \\
-- & -- & -- & -- & 3.0 & 30.4 & $1.7 \times 10^{-10}$ & 30  & 629  \\
NGC 1068 & $40.67^\circ$ & $-0.01^\circ$ & 0.50 (--) & 2.5 & 22.3 & $3.0 \times 10^{-10}$ & 39  & 1027  \\
-- & -- & -- & -- & 3.0 & 30.4 & $1.8 \times 10^{-10}$ & 31  & 631  \\
Mrk 501 & $253.47^\circ$ & $39.76^\circ$ & 0.51 (--) & 2.5 & 15.8 & $7.5 \times 10^{-11}$ & 47  & 719  \\
-- & -- & -- & -- & 3.0 & 23.1 & $7.1 \times 10^{-11}$ & 34  & 506  \\
NGC 2110 & $88.05^\circ$ & $-7.46^\circ$ & 0.52 (--) & 2.5 & 21.4 & $4.0 \times 10^{-10}$ & 38  & 1107  \\
-- & -- & -- & -- & 3.0 & 31.7 & $2.6 \times 10^{-10}$ & 30  & 670  \\
Mrk 421 & $166.11^\circ$ & $38.21^\circ$ & 0.52 (--) & 2.5 & 15.1 & $7.2 \times 10^{-11}$ & 47  & 756  \\
-- & -- & -- & -- & 3.0 & 23.1 & $7.1 \times 10^{-11}$ & 34  & 516  \\
NGC 4102 & $181.60^\circ$ & $52.71^\circ$ & 0.52 (--) & 2.5 & 16.3 & $8.8 \times 10^{-11}$ & 50  & 685  \\
-- & -- & -- & -- & 3.0 & 24.1 & $8.1 \times 10^{-11}$ & 36  & 474  \\
NGC 1142 & $43.80^\circ$ & $-0.18^\circ$ & 0.52 (--) & 2.5 & 22.2 & $3.1 \times 10^{-10}$ & 39  & 1032  \\
-- & -- & -- & -- & 3.0 & 30.3 & $1.8 \times 10^{-10}$ & 31  & 630  \\
3C 111 & $64.59^\circ$ & $38.03^\circ$ & 0.56 (--) & 2.5 & 15.0 & $7.1 \times 10^{-11}$ & 47  & 758  \\
-- & -- & -- & -- & 3.0 & 23.3 & $7.2 \times 10^{-11}$ & 34  & 511  \\
NGC 3079 & $150.49^\circ$ & $55.68^\circ$ & 0.56 (--) & 2.5 & 17.6 & $1.0 \times 10^{-10}$ & 51  & 626  \\
-- & -- & -- & -- & 3.0 & 24.7 & $8.8 \times 10^{-11}$ & 37  & 459  \\
IGR J21277+5656 & $321.94^\circ$ & $56.94^\circ$ & 0.59 (--) & 2.5 & 18.3 & $1.1 \times 10^{-10}$ & 51  & 659  \\
-- & -- & -- & -- & 3.0 & 24.1 & $8.8 \times 10^{-11}$ & 38  & 467  \\
Cygnus A & $299.87^\circ$ & $40.73^\circ$ & 0.70 (--) & 2.5 & 16.3 & $7.7 \times 10^{-11}$ & 48  & 739  \\
-- & -- & -- & -- & 3.0 & 23.5 & $7.2 \times 10^{-11}$ & 34  & 509  \\
NGC 4388 & $186.44^\circ$ & $12.66^\circ$ & 1.00 (--) & 2.5 & 20.7 & $1.5 \times 10^{-10}$ & 43  & 927  \\
-- & -- & -- & -- & 3.0 & 28.7 & $1.2 \times 10^{-10}$ & 32  & 589  \\
3C 452 & $341.45^\circ$ & $39.69^\circ$ & 1.00 (--) & 2.5 & 15.6 & $7.4 \times 10^{-11}$ & 47  & 705  \\
-- & -- & -- & -- & 3.0 & 23.3 & $7.2 \times 10^{-11}$ & 34  & 505  \\
UGC 3752 & $108.52^\circ$ & $35.28^\circ$ & 1.00 (--) & 2.5 & 15.8 & $7.6 \times 10^{-11}$ & 47  & 817  \\
-- & -- & -- & -- & 3.0 & 23.5 & $7.3 \times 10^{-11}$ & 33  & 549  \\
NGC 6240 & $253.25^\circ$ & $2.40^\circ$ & 1.00 (--) & 2.5 & 22.3 & $2.8 \times 10^{-10}$ & 38  & 965  \\
-- & -- & -- & -- & 3.0 & 30.4 & $1.7 \times 10^{-10}$ & 31  & 620  \\
3C 120 & $68.30^\circ$ & $5.35^\circ$ & 1.00 (--) & 2.5 & 22.9 & $2.5 \times 10^{-10}$ & 39  & 964  \\
-- & -- & -- & -- & 3.0 & 30.2 & $1.5 \times 10^{-10}$ & 31  & 602  \\
NGC 5252 & $204.57^\circ$ & $4.54^\circ$ & 1.00 (--) & 2.5 & 22.2 & $2.5 \times 10^{-10}$ & 39  & 1020  \\
-- & -- & -- & -- & 3.0 & 30.0 & $1.6 \times 10^{-10}$ & 31  & 618  \\
Mrk 79 & $115.64^\circ$ & $49.81^\circ$ & 1.00 (--) & 2.5 & 15.8 & $8.1 \times 10^{-11}$ & 50  & 700  \\
-- & -- & -- & -- & 3.0 & 23.3 & $7.6 \times 10^{-11}$ & 36  & 484  \\
ESO 138-1 & $252.83^\circ$ & $-59.23^\circ$ & 1.00 (--) & 2.5 & 10.9 & $3.3 \times 10^{-9}$ & 28  & 570  \\
-- & -- & -- & -- & 3.0 & 14.4 & $1.8 \times 10^{-9}$ & 25  & 445  \\
NGC 4051 & $180.79^\circ$ & $44.53^\circ$ & 1.00 (--) & 2.5 & 16.0 & $7.8 \times 10^{-11}$ & 48  & 718  \\
-- & -- & -- & -- & 3.0 & 23.9 & $7.5 \times 10^{-11}$ & 35  & 515  \\
Mrk 110 & $141.30^\circ$ & $52.29^\circ$ & 1.00 (--) & 2.5 & 17.3 & $9.3 \times 10^{-11}$ & 50  & 682  \\
-- & -- & -- & -- & 3.0 & 24.6 & $8.3 \times 10^{-11}$ & 36  & 472  \\
LEDA 168563 & $73.02^\circ$ & $49.55^\circ$ & 1.00 (--) & 2.5 & 15.5 & $7.9 \times 10^{-11}$ & 50  & 702  \\
-- & -- & -- & -- & 3.0 & 23.0 & $7.5 \times 10^{-11}$ & 36  & 481  \\
NGC 3227 & $155.88^\circ$ & $19.87^\circ$ & 1.00 (--) & 2.5 & 17.8 & $1.1 \times 10^{-10}$ & 45  & 939  \\
-- & -- & -- & -- & 3.0 & 25.8 & $9.3 \times 10^{-11}$ & 32  & 566  \\
3C 454.3 & $343.49^\circ$ & $16.15^\circ$ & 1.00 (--) & 2.5 & 18.7 & $1.2 \times 10^{-10}$ & 45  & 893  \\
-- & -- & -- & -- & 3.0 & 27.4 & $1.1 \times 10^{-10}$ & 33  & 566  \\
NGC 5548 & $214.50^\circ$ & $25.14^\circ$ & 1.00 (--) & 2.5 & 16.3 & $8.9 \times 10^{-11}$ & 45  & 797  \\
-- & -- & -- & -- & 3.0 & 24.6 & $8.3 \times 10^{-11}$ & 31  & 556  \\
2MASX J20145928 & $303.75^\circ$ & $25.38^\circ$ & 1.00 (--) & 2.5 & 16.4 & $9.0 \times 10^{-11}$ & 45  & 804  \\
+2523010 & ~ & ~ & ~ & ~ & ~ & ~ & ~ & ~ \\
-- & -- & -- & -- & 3.0 & 24.9 & $8.4 \times 10^{-11}$ & 31  & 550  \\
3C 382 & $278.76^\circ$ & $32.70^\circ$ & 1.00 (--) & 2.5 & 14.9 & $7.3 \times 10^{-11}$ & 46  & 746  \\
-- & -- & -- & -- & 3.0 & 23.5 & $7.4 \times 10^{-11}$ & 33  & 545  \\
LEDA 86269 & $71.04^\circ$ & $28.22^\circ$ & 1.00 (--) & 2.5 & 15.3 & $7.9 \times 10^{-11}$ & 45  & 863  \\
-- & -- & -- & -- & 3.0 & 24.1 & $7.9 \times 10^{-11}$ & 32  & 545  \\
Cen A & $201.37^\circ$ & $-43.02^\circ$ & 1.00 (--) & 2.5 & 16.1 & $4.1 \times 10^{-9}$ & 31  & 1470  \\
-- & -- & -- & -- & 3.0 & 26.8 & $2.9 \times 10^{-9}$ & 23  & 667  \\
\bottomrule
\end{longtable}

\begin{longtable}{lllllllll}
\caption{\label{tab:catapbresults}
Summary of the catalog search done on a selection of 56 Galactic sources. Sources are ordered in order of descending local significance. For each source, we list the equatorial coordinates (J2000 equinox) converted from the Galactic coordinates listed in BASS DR1 catalog, the local (global) p-value, the spectral index for the corresponding upper limit, the 90\% CL upper limits on the number of source neutrinos as well as the astrophysical $\nu + \bar{\nu}$ per-flavor flux evaluated at 1 TeV with units of TeV$^{-1}$ cm$^{-2}$ s$^{-1}$, and the energies that bracket the 95\% central energy range for each respective limit in GeV. For each candidate name, we list the associated source as in \citealt{4FGL}, if applicable.
}
\\
\toprule
Candidate name & $\alpha$ & $\delta$ & $p_\mathrm{local}$~($p_\mathrm{global}$) & $\gamma$ & $n_{\mathrm{s},90\%}$ & $\Phi |_{1\mathrm{TeV}}$ & $E_{2.5\%}$ & $E_{97.5\%}$ \\
\midrule
LS I +61 303 & $40.14^\circ$ & $61.23^\circ$ & 0.01 (0.30) & 2.5 & 51.9 & $3.4 \times 10^{-10}$ & 50  & 649  \\
-- & -- & -- & -- & 3.0 & 70.1 & $2.7 \times 10^{-10}$ & 38  & 463  \\
4FGL J0340.4+5302 & $55.11^\circ$ & $53.04^\circ$ & 0.03 (--) & 2.5 & 45.4 & $2.5 \times 10^{-10}$ & 50  & 682  \\
-- & -- & -- & -- & 3.0 & 64.5 & $2.2 \times 10^{-10}$ & 36  & 470  \\
SNR G054.4-00.3 & $293.59^\circ$ & $19.00^\circ$ & 0.04 (--) & 2.5 & 45.9 & $2.8 \times 10^{-10}$ & 45  & 885  \\
-- & -- & -- & -- & 3.0 & 67.0 & $2.5 \times 10^{-10}$ & 32  & 564  \\
4FGL J2108.0+5155 & $317.03^\circ$ & $51.93^\circ$ & 0.09 (--) & 2.5 & 36.9 & $2.0 \times 10^{-10}$ & 51  & 683  \\
-- & -- & -- & -- & 3.0 & 52.6 & $1.8 \times 10^{-10}$ & 36  & 473  \\
4FGL J0620.4+1445 & $95.10^\circ$ & $14.76^\circ$ & 0.10 (--) & 2.5 & 39.1 & $2.7 \times 10^{-10}$ & 44  & 900  \\
-- & -- & -- & -- & 3.0 & 57.0 & $2.3 \times 10^{-10}$ & 32  & 601  \\
4FGL J0330.7+5845 & $52.69^\circ$ & $58.75^\circ$ & 0.10 (--) & 2.5 & 38.3 & $2.4 \times 10^{-10}$ & 51  & 665  \\
-- & -- & -- & -- & 3.0 & 51.5 & $1.9 \times 10^{-10}$ & 38  & 462  \\
4FGL J1931.1+1656 & $292.78^\circ$ & $16.95^\circ$ & 0.12 (--) & 2.5 & 36.7 & $2.4 \times 10^{-10}$ & 44  & 903  \\
-- & -- & -- & -- & 3.0 & 51.9 & $2.0 \times 10^{-10}$ & 33  & 566  \\
4FGL J1008.1-5706c & $152.03^\circ$ & $-57.10^\circ$ & 0.13 (--) & 2.5 & 21.0 & $6.6 \times 10^{-9}$ & 30  & 531  \\
-- & -- & -- & -- & 3.0 & 28.7 & $3.5 \times 10^{-9}$ & 25  & 441  \\
4FGL J2038.4+4212 & $309.63^\circ$ & $42.21^\circ$ & 0.21 (--) & 2.5 & 26.1 & $1.2 \times 10^{-10}$ & 48  & 731  \\
-- & -- & -- & -- & 3.0 & 38.4 & $1.2 \times 10^{-10}$ & 35  & 522  \\
W 28 & $270.34^\circ$ & $-23.44^\circ$ & 0.24 (--) & 2.5 & 46.9 & $3.2 \times 10^{-9}$ & 32  & 1539  \\
-- & -- & -- & -- & 3.0 & 78.3 & $2.3 \times 10^{-9}$ & 28  & 789  \\
Sim 147 & $85.10^\circ$ & $27.94^\circ$ & 0.26 (--) & 2.5 & 23.6 & $1.2 \times 10^{-10}$ & 45  & 848  \\
-- & -- & -- & -- & 3.0 & 37.1 & $1.2 \times 10^{-10}$ & 32  & 545  \\
Monoceros & $99.86^\circ$ & $6.93^\circ$ & 0.26 (--) & 2.5 & 33.1 & $3.3 \times 10^{-10}$ & 41  & 959  \\
-- & -- & -- & -- & 3.0 & 44.5 & $2.2 \times 10^{-10}$ & 33  & 624  \\
HESS J1632-478 & $248.25^\circ$ & $-47.77^\circ$ & 0.27 (--) & 2.5 & 19.6 & $5.4 \times 10^{-9}$ & 32  & 1120  \\
-- & -- & -- & -- & 3.0 & 32.3 & $3.6 \times 10^{-9}$ & 26  & 675  \\
IC 447 & $94.31^\circ$ & $22.58^\circ$ & 0.28 (--) & 2.5 & 23.4 & $1.3 \times 10^{-10}$ & 45  & 825  \\
-- & -- & -- & -- & 3.0 & 36.8 & $1.3 \times 10^{-10}$ & 32  & 558  \\
HBH 9 & $75.08^\circ$ & $46.66^\circ$ & 0.28 (--) & 2.5 & 23.4 & $1.2 \times 10^{-10}$ & 49  & 691  \\
-- & -- & -- & -- & 3.0 & 34.0 & $1.1 \times 10^{-10}$ & 34  & 497  \\
MSH 15-56 & $154.74^\circ$ & $-58.94^\circ$ & 0.31 (--) & 2.5 & 15.2 & $4.7 \times 10^{-9}$ & 28  & 572  \\
-- & -- & -- & -- & 3.0 & 19.7 & $2.5 \times 10^{-9}$ & 25  & 439  \\
4FGL J1552.9-5607e & $238.26^\circ$ & $-56.13^\circ$ & 0.32 (--) & 2.5 & 15.3 & $4.9 \times 10^{-9}$ & 29  & 557  \\
-- & -- & -- & -- & 3.0 & 20.4 & $2.5 \times 10^{-9}$ & 24  & 421  \\
4FGL J1906.9+0712 & $286.75^\circ$ & $7.21^\circ$ & 0.37 (--) & 2.5 & 26.9 & $2.7 \times 10^{-10}$ & 41  & 976  \\
-- & -- & -- & -- & 3.0 & 37.4 & $1.8 \times 10^{-10}$ & 32  & 636  \\
4FGL J1742.8-2246 & $265.73^\circ$ & $-22.77^\circ$ & 0.38 (--) & 2.5 & 37.6 & $2.4 \times 10^{-9}$ & 33  & 1451  \\
-- & -- & -- & -- & 3.0 & 62.1 & $1.7 \times 10^{-9}$ & 27  & 763  \\
W 51C & $290.82^\circ$ & $14.15^\circ$ & 0.41 (--) & 2.5 & 23.3 & $1.6 \times 10^{-10}$ & 44  & 937  \\
-- & -- & -- & -- & 3.0 & 32.8 & $1.3 \times 10^{-10}$ & 33  & 610  \\
4FGL J1608.8-4803 & $242.23^\circ$ & $-48.06^\circ$ & 0.50 (--) & 2.5 & 13.3 & $3.8 \times 10^{-9}$ & 32  & 1151  \\
-- & -- & -- & -- & 3.0 & 21.1 & $2.5 \times 10^{-9}$ & 26  & 679  \\
4FGL J0426.5+5434 & $66.63^\circ$ & $54.57^\circ$ & 0.55 (--) & 2.5 & 16.6 & $9.4 \times 10^{-11}$ & 50  & 674  \\
-- & -- & -- & -- & 3.0 & 23.9 & $8.4 \times 10^{-11}$ & 37  & 468  \\
SNR G329.7+00.4 & $240.34^\circ$ & $-52.40^\circ$ & 0.57 (--) & 2.5 & 12.2 & $3.6 \times 10^{-9}$ & 31  & 1089  \\
-- & -- & -- & -- & 3.0 & 17.7 & $2.1 \times 10^{-9}$ & 25  & 480  \\
HESS J1813-178 & $273.29^\circ$ & $-17.62^\circ$ & 0.59 (--) & 2.5 & 30.4 & $1.3 \times 10^{-9}$ & 34  & 1347  \\
-- & -- & -- & -- & 3.0 & 46.8 & $8.2 \times 10^{-10}$ & 28  & 779  \\
Cygnus X & $307.17^\circ$ & $41.17^\circ$ & 0.60 (--) & 2.5 & 15.9 & $7.6 \times 10^{-11}$ & 48  & 744  \\
-- & -- & -- & -- & 3.0 & 23.4 & $7.2 \times 10^{-11}$ & 35  & 515  \\
4FGL J1808.2-1055 & $272.06^\circ$ & $-10.92^\circ$ & 0.61 (--) & 2.5 & 25.1 & $5.9 \times 10^{-10}$ & 37  & 1105  \\
-- & -- & -- & -- & 3.0 & 35.4 & $3.5 \times 10^{-10}$ & 30  & 659  \\
4FGL J0709.1-1034c & $107.29^\circ$ & $-10.57^\circ$ & 0.61 (--) & 2.5 & 25.2 & $5.7 \times 10^{-10}$ & 37  & 1125  \\
-- & -- & -- & -- & 3.0 & 35.6 & $3.4 \times 10^{-10}$ & 30  & 658  \\
4FGL J1855.2+0456 & $283.82^\circ$ & $4.94^\circ$ & 0.61 (--) & 2.5 & 22.0 & $2.4 \times 10^{-10}$ & 39  & 1033  \\
-- & -- & -- & -- & 3.0 & 29.8 & $1.5 \times 10^{-10}$ & 31  & 613  \\
4FGL J1908.7+0812 & $287.19^\circ$ & $8.20^\circ$ & 0.63 (--) & 2.5 & 22.0 & $2.1 \times 10^{-10}$ & 41  & 990  \\
-- & -- & -- & -- & 3.0 & 29.7 & $1.4 \times 10^{-10}$ & 32  & 630  \\
Rosette & $98.57^\circ$ & $4.61^\circ$ & 1.00 (--) & 2.5 & 22.3 & $2.5 \times 10^{-10}$ & 39  & 1024  \\
-- & -- & -- & -- & 3.0 & 29.7 & $1.5 \times 10^{-10}$ & 31  & 616  \\
4FGL J0609.0+2006 & $92.26^\circ$ & $20.10^\circ$ & 1.00 (--) & 2.5 & 17.5 & $1.0 \times 10^{-10}$ & 44  & 939  \\
-- & -- & -- & -- & 3.0 & 26.1 & $9.3 \times 10^{-11}$ & 32  & 571  \\
HESS J1857+026 & $284.45^\circ$ & $2.78^\circ$ & 1.00 (--) & 2.5 & 22.2 & $2.7 \times 10^{-10}$ & 39  & 988  \\
-- & -- & -- & -- & 3.0 & 30.5 & $1.7 \times 10^{-10}$ & 31  & 626  \\
$\gamma$ Cygni & $305.27^\circ$ & $40.52^\circ$ & 1.00 (--) & 2.5 & 16.0 & $7.6 \times 10^{-11}$ & 48  & 738  \\
-- & -- & -- & -- & 3.0 & 23.1 & $7.1 \times 10^{-11}$ & 34  & 507  \\
4FGL J2032.6+4053 & $308.15^\circ$ & $40.89^\circ$ & 1.00 (--) & 2.5 & 16.0 & $7.6 \times 10^{-11}$ & 48  & 736  \\
-- & -- & -- & -- & 3.0 & 23.3 & $7.2 \times 10^{-11}$ & 35  & 510  \\
4FGL J2045.2+5026e & $311.32^\circ$ & $50.44^\circ$ & 1.00 (--) & 2.5 & 16.1 & $8.3 \times 10^{-11}$ & 50  & 682  \\
-- & -- & -- & -- & 3.0 & 23.5 & $7.7 \times 10^{-11}$ & 36  & 477  \\
W 44 & $283.99^\circ$ & $1.35^\circ$ & 1.00 (--) & 2.5 & 22.0 & $2.9 \times 10^{-10}$ & 38  & 985  \\
-- & -- & -- & -- & 3.0 & 29.8 & $1.7 \times 10^{-10}$ & 31  & 618  \\
W 49B & $287.77^\circ$ & $9.09^\circ$ & 1.00 (--) & 2.5 & 21.4 & $1.9 \times 10^{-10}$ & 41  & 996  \\
-- & -- & -- & -- & 3.0 & 29.8 & $1.4 \times 10^{-10}$ & 32  & 632  \\
Kes 79 & $283.10^\circ$ & $0.63^\circ$ & 1.00 (--) & 2.5 & 22.2 & $3.0 \times 10^{-10}$ & 38  & 1021  \\
-- & -- & -- & -- & 3.0 & 30.5 & $1.8 \times 10^{-10}$ & 31  & 629  \\
4FGL J0850.8-4239 & $132.71^\circ$ & $-42.66^\circ$ & 1.00 (--) & 2.5 & 15.5 & $3.9 \times 10^{-9}$ & 31  & 1155  \\
-- & -- & -- & -- & 3.0 & 26.7 & $2.8 \times 10^{-9}$ & 25  & 687  \\
4FGL J1814.2-1012 & $273.56^\circ$ & $-10.21^\circ$ & 1.00 (--) & 2.5 & 24.7 & $5.5 \times 10^{-10}$ & 37  & 1094  \\
-- & -- & -- & -- & 3.0 & 34.9 & $3.3 \times 10^{-10}$ & 30  & 635  \\
Eta Carinae & $161.28^\circ$ & $-59.68^\circ$ & 1.00 (--) & 2.5 & 10.6 & $3.2 \times 10^{-9}$ & 28  & 568  \\
-- & -- & -- & -- & 3.0 & 13.7 & $1.7 \times 10^{-9}$ & 25  & 452  \\
4FGL J1351.6-6142 & $207.90^\circ$ & $-61.70^\circ$ & 1.00 (--) & 2.5 & 9.6 & $2.9 \times 10^{-9}$ & 29  & 512  \\
-- & -- & -- & -- & 3.0 & 12.8 & $1.6 \times 10^{-9}$ & 26  & 437  \\
PSR J1358-6025 & $209.60^\circ$ & $-60.45^\circ$ & 1.00 (--) & 2.5 & 10.1 & $3.1 \times 10^{-9}$ & 29  & 562  \\
-- & -- & -- & -- & 3.0 & 13.2 & $1.7 \times 10^{-9}$ & 24  & 453  \\
4FGL J1405.1-6119 & $211.30^\circ$ & $-61.34^\circ$ & 1.00 (--) & 2.5 & 9.7 & $3.0 \times 10^{-9}$ & 29  & 550  \\
-- & -- & -- & -- & 3.0 & 12.8 & $1.6 \times 10^{-9}$ & 26  & 451  \\
SNR G316.3-00.0 & $220.57^\circ$ & $-60.08^\circ$ & 1.00 (--) & 2.5 & 10.5 & $3.2 \times 10^{-9}$ & 28  & 568  \\
-- & -- & -- & -- & 3.0 & 13.5 & $1.7 \times 10^{-9}$ & 24  & 455  \\
PSR J1447-5757 & $221.85^\circ$ & $-57.96^\circ$ & 1.00 (--) & 2.5 & 11.3 & $3.6 \times 10^{-9}$ & 31  & 578  \\
-- & -- & -- & -- & 3.0 & 14.9 & $1.9 \times 10^{-9}$ & 26  & 455  \\
NVSS J183922-055321 & $280.13^\circ$ & $-6.02^\circ$ & 1.00 (--) & 2.5 & 20.9 & $3.7 \times 10^{-10}$ & 38  & 1114  \\
-- & -- & -- & -- & 3.0 & 30.8 & $2.4 \times 10^{-10}$ & 31  & 669  \\
MSH 15-52 & $228.58^\circ$ & $-59.16^\circ$ & 1.00 (--) & 2.5 & 11.4 & $3.5 \times 10^{-9}$ & 28  & 571  \\
-- & -- & -- & -- & 3.0 & 13.9 & $1.7 \times 10^{-9}$ & 25  & 445  \\
4FGL J1547.5-5130 & $236.88^\circ$ & $-51.51^\circ$ & 1.00 (--) & 2.5 & 11.8 & $3.4 \times 10^{-9}$ & 31  & 1209  \\
-- & -- & -- & -- & 3.0 & 17.6 & $2.0 \times 10^{-9}$ & 24  & 487  \\
4FGL J2056.4+4351c & $314.12^\circ$ & $43.85^\circ$ & 1.00 (--) & 2.5 & 15.8 & $7.7 \times 10^{-11}$ & 47  & 704  \\
-- & -- & -- & -- & 3.0 & 23.3 & $7.3 \times 10^{-11}$ & 35  & 514  \\
4FGL J1626.6-4251 & $246.65^\circ$ & $-42.86^\circ$ & 1.00 (--) & 2.5 & 15.7 & $4.0 \times 10^{-9}$ & 31  & 1461  \\
-- & -- & -- & -- & 3.0 & 26.2 & $2.8 \times 10^{-9}$ & 24  & 687  \\
4FGL J0844.1-4330 & $131.03^\circ$ & $-43.51^\circ$ & 1.00 (--) & 2.5 & 16.3 & $4.1 \times 10^{-9}$ & 30  & 1338  \\
-- & -- & -- & -- & 3.0 & 26.9 & $2.9 \times 10^{-9}$ & 23  & 689  \\
4FGL J1812.2-0856 & $273.08^\circ$ & $-8.95^\circ$ & 1.00 (--) & 2.5 & 23.3 & $4.8 \times 10^{-10}$ & 37  & 985  \\
-- & -- & -- & -- & 3.0 & 33.7 & $3.0 \times 10^{-10}$ & 30  & 635  \\
4FGL J0904.7-4908c & $136.17^\circ$ & $-49.14^\circ$ & 1.00 (--) & 2.5 & 12.2 & $3.6 \times 10^{-9}$ & 32  & 1033  \\
-- & -- & -- & -- & 3.0 & 19.3 & $2.3 \times 10^{-9}$ & 25  & 646  \\
4FGL J1534.0-5232 & $233.50^\circ$ & $-52.55^\circ$ & 1.00 (--) & 2.5 & 12.4 & $3.7 \times 10^{-9}$ & 31  & 1088  \\
-- & -- & -- & -- & 3.0 & 18.0 & $2.2 \times 10^{-9}$ & 25  & 475  \\
W 3 & $35.63^\circ$ & $61.94^\circ$ & 1.00 (--) & 2.5 & 20.3 & $1.3 \times 10^{-10}$ & 50  & 656  \\
-- & -- & -- & -- & 3.0 & 27.3 & $1.1 \times 10^{-10}$ & 38  & 465  \\
\bottomrule
\end{longtable}

\bibliography{references}{}

@article{IceCube:2013Science,
  author        = {{Aartsen}, M.~G. and others},
  title         = {{Evidence for High-Energy Extraterrestrial Neutrinos at the IceCube Detector}},
  journal       = {Science},
  year          = 2013,
  month         = nov,
  volume        = {342},
  number        = {6161},
  eid           = {1242856},
  pages         = {1242856},
  doi           = {10.1126/science.1242856},
  archiveprefix = {arXiv},
  eprint        = {1311.5238},
  primaryclass  = {astro-ph.HE},
  adsurl        = {https://ui.adsabs.harvard.edu/abs/2013Sci...342E...1I}
}

@article{Icecube:2016Detector,
      author         = "{Aartsen}, M.~G. and others",
      title          = "{The IceCube Neutrino Observatory: Instrumentation and
                        Online Systems}",
      collaboration  = "IceCube",
      journal        = "JINST",
      volume         = "12",
      year           = "2017",
      number         = "03",
      pages          = "P03012",
      doi            = "10.1088/1748-0221/12/03/P03012",
      eprint         = "1612.05093",
      archivePrefix  = "arXiv",
      primaryClass   = "astro-ph.IM",
      SLACcitation   = "%%CITATION = ARXIV:1612.05093;%%"
}

@article{IceCube:2018ScienceAlert,
  author        = {{Aartsen}, M.~G. and others},
  title         = {{Multimessenger observations of a flaring blazar coincident with high-energy neutrino IceCube-170922A}},
  journal       = {Science},
  year          = {2018},
  month         = jul,
  volume        = {361},
  number        = {6398},
  eid           = {eaat1378},
  pages         = {eaat1378},
  doi           = {10.1126/science.aat1378},
  archiveprefix = {arXiv},
  eprint        = {1807.08816},
  primaryclass  = {astro-ph.HE},
  adsurl        = {https://ui.adsabs.harvard.edu/abs/2018Sci...361.1378I}
}

@article{IceCube:2018ScienceFlare,
  author        = {{Aartsen}, M.~G. and others},
  title         = {{Neutrino emission from the direction of the blazar TXS 0506+056 prior to the IceCube-170922A alert}},
  journal       = {Science},
  year          = {2018},
  month         = jul,
  volume        = {361},
  number        = {6398},
  pages         = {147-151},
  doi           = {10.1126/science.aat2890},
  archiveprefix = {arXiv},
  eprint        = {1807.08794},
  primaryclass  = {astro-ph.HE},
  adsurl        = {https://ui.adsabs.harvard.edu/abs/2018Sci...361..147I}
}

@article{IceCube:2022Science,
  author        = {{Abbasi}, R. and others},
  title         = {{Evidence for neutrino emission from the nearby active galaxy NGC 1068}},
  journal       = {Science},
  year          = 2022,
  month         = nov,
  volume        = {378},
  number        = {6619},
  pages         = {538-543},
  doi           = {10.1126/science.abg3395},
  archiveprefix = {arXiv},
  eprint        = {2211.09972},
  primaryclass  = {astro-ph.HE},
  adsurl        = {https://ui.adsabs.harvard.edu/abs/2022Sci...378..538I}
}

@article{abbasi2025evidenceneutrinoemissionxray,
doi = {10.3847/2041-8213/ae4aad},
url = {https://doi.org/10.3847/2041-8213/ae4aad},
year = {2026},
month = {mar},
publisher = {The American Astronomical Society},
volume = {1000},
number = {1},
pages = {L26},
author = {Abbasi, R. and others},
title = {Evidence for Neutrino Emission from X-Ray-bright Active Galactic Nuclei with IceCube},
journal = {\apjl},
}

@article{Abbasi2026EvidenceSouth,
	author = {Abbasi, R. and others},
	doi = {10.3847/2041-8213/ae4aac},
	issn = {2041-8205},
	number = {2},
	year = {2026},
	month = {mar 23},
	pages = {L37},
	publisher = {American Astronomical Society},
	title = {Evidence for {Neutrino} {Emission} from {X}-{Ray} {Bright} {Seyfert} {Galaxies} in the {Southern} {Hemisphere} {Using} {Enhanced} {Starting} {Track} {Events} with {IceCube}},
	url = {http://dx.doi.org/10.3847/2041-8213/ae4aac},
	volume = {1000},
    journal = {\apjl},
}

@article{Abbasi2026Physics,
	author = {Abbasi, R. and others},
	journal = {Physical Review D},
	doi = {10.1103/nnjw-jp1n},
	issn = {2470-0010},
	number = {7},
	year = {2026},
	month = {apr 15},
	publisher = {American Physical Society (APS)},
	title = {Physics potential of the {IceCube} {Upgrade} for atmospheric neutrino oscillations},
	url = {http://dx.doi.org/10.1103/nnjw-jp1n},
	volume = {113},
}

@article{Murase2016Hidden,
	author = {Murase, Kohta and Guetta, Dafne and Ahlers, Markus},
	journal = {Physical Review Letters},
	doi = {10.1103/physrevlett.116.071101},
	issn = {0031-9007},
	number = {7},
	year = {2016},
	month = {feb 18},
	publisher = {American Physical Society (APS)},
	title = {Hidden {Cosmic}-{Ray} {Accelerators} as an {Origin} of {TeV}-{PeV} {Cosmic} {Neutrinos}},
	url = {http://dx.doi.org/10.1103/PhysRevLett.116.071101},
	volume = {116},
}

@article{IceCube:2023Science,
  author        = {{Abbasi}, R. and others},
  title         = {{Observation of high-energy neutrinos from the Galactic plane}},
  journal       = {Science},
  year          = 2023,
  month         = jun,
  volume        = {380},
  number        = {6652},
  pages         = {1338-1343},
  doi           = {10.1126/science.adc9818},
  archiveprefix = {arXiv},
  eprint        = {2307.04427},
  primaryclass  = {astro-ph.HE},
  adsurl        = {https://ui.adsabs.harvard.edu/abs/2023Sci...380.1338I}
}

@article{AHLERS201873,
title = {Opening a new window onto the universe with IceCube},
journal = {Progress in Particle and Nuclear Physics},
volume = {102},
pages = {73-88},
year = {2018},
issn = {0146-6410},
doi = {https://doi.org/10.1016/j.ppnp.2018.05.001},
url = {https://www.sciencedirect.com/science/article/pii/S0146641018300346},
author = {Ahlers, M. and Halzen, F.},
}

@article{ABBASI2012615,
title = {The design and performance of IceCube DeepCore},
journal = {Astroparticle Physics},
volume = {35},
number = {10},
pages = {615-624},
year = {2012},
issn = {0927-6505},
doi = {https://doi.org/10.1016/j.astropartphys.2012.01.004},
url = {https://www.sciencedirect.com/science/article/pii/S0927650512000254},
author        = {{Abbasi}, R. and others},
}

@article{Abbasi2024Search,
	author = {Abbasi, R. and others},
	journal = {\apj},
	doi = {10.3847/1538-4357/ad220b},
	issn = {0004-637X},
	number = {2},
	year = {2024},
	month = {mar 22},
	pages = {126},
	publisher = {American Astronomical Society},
	title = {Search for 10--1000 {GeV} {Neutrinos} from {Gamma}-{Ray} {Bursts} with {IceCube}},
	url = {http://dx.doi.org/10.3847/1538-4357/ad220b},
	volume = {964},
}

@article{Abbasi2023Limits,
	author = {Abbasi, R. and others},
	journal = {\apjl},
	doi = {10.3847/2041-8213/acc077},
	issn = {2041-8205},
	number = {1},
	year = {2023},
	month = {mar 1},
	pages = {L26},
	publisher = {American Astronomical Society},
	title = {Limits on {Neutrino} {Emission} from {GRB} 221009A from {MeV} to {PeV} {Using} the {IceCube} {Neutrino} {Observatory}},
	url = {http://dx.doi.org/10.3847/2041-8213/acc077},
	volume = {946},
}

@article{Abbasi2023Search,
	author = {Abbasi, R. and others},
	journal = {\apj},
	doi = {10.3847/1538-4357/acdc1b},
	issn = {0004-637X},
	number = {2},
	year = {2023},
	month = {aug 1},
	pages = {160},
	publisher = {American Astronomical Society},
	title = {Search for sub-{TeV} {Neutrino} {Emission} from {Novae} with {IceCube}-{DeepCore}},
	url = {http://dx.doi.org/10.3847/1538-4357/acdc1b},
	volume = {953},
}

@article{Abbasi2023SearchGW,
	author = {Abbasi, R. and others},
	journal = {\apj},
	doi = {10.3847/1538-4357/aceefc},
	issn = {0004-637X},
	number = {2},
	year = {2023},
	month = {dec 1},
	pages = {96},
	publisher = {American Astronomical Society},
	title = {A {Search} for {IceCube} {Sub}-{TeV} {Neutrinos} {Correlated} with {Gravitational}-wave {Events} {Detected} {By} {LIGO}/{Virgo}},
	url = {http://dx.doi.org/10.3847/1538-4357/aceefc},
	volume = {959},
}

@article{Ricci:2017ApJS,
    author = {{Ricci}, C. and {Trakhtenbrot}, B. and {Koss}, M.~J. and {Ueda}, Y. and {Delvecchio}, I. and {Treister}, E. and {Schawinski}, K. and {Paltani}, S. and {Oh}, K. and {Lamperti}, I. and {Berney}, S. and {Gandhi}, P. and {Ichikawa}, K. and {Bauer}, F.~E. and {Ho}, L.~C. and {Asmus}, D. and {Beckmann}, V. and {Soldi}, S. and {Balokovi{\'c}}, M. and {Gehrels}, N. and {Markwardt}, C.~B.},
    title = "{BAT AGN Spectroscopic Survey. V. X-Ray Properties of the Swift/BAT 70-month AGN Catalog}",
    journal = {\apjs},
    year = 2017,
    month = dec,
    volume = {233},
    number = {2},
    eid = {17},
    pages = {17},
    doi = {10.3847/1538-4365/aa96ad},
    archivePrefix = {arXiv},
    eprint = {1709.03989},
    primaryClass = {astro-ph.HE},
    adsurl = {https://ui.adsabs.harvard.edu/abs/2017ApJS..233...17R}
}

@article{Abdollahi2022Search,
	author = {Abdollahi, S. and others},
	journal = {\apj},
	doi = {10.3847/1538-4357/ac704f},
	issn = {0004-637X},
	number = {2},
	year = {2022},
	month = {jul 1},
	pages = {204},
	publisher = {American Astronomical Society},
	title = {Search for {New} {Cosmic}-{Ray} {Acceleration} {Sites} within the 4FGL {Catalog} {Galactic} {Plane} {Sources}},
	url = {http://dx.doi.org/10.3847/1538-4357/ac704f},
	volume = {933},
}

@article{1FGL,
  author        = {{Abdo}, A.~A. and others},
  title         = {{Fermi Large Area Telescope First Source Catalog}},
  journal       = {\apjs},
  year          = 2010,
  month         = jun,
  volume        = {188},
  number        = {2},
  pages         = {405-436},
  doi           = {10.1088/0067-0049/188/2/405},
  archiveprefix = {arXiv},
  eprint        = {1002.2280},
  primaryclass  = {astro-ph.HE},
  adsurl        = {https://ui.adsabs.harvard.edu/abs/2010ApJS..188..405A}
}

@article{Eichmann:2022ApJ,
  author        = {{Eichmann}, Bj{\"o}rn and {Oikonomou}, Foteini and {Salvatore}, Silvia and {Dettmar}, Ralf-J{\"u}rgen and {Tjus}, Julia Becker},
  title         = {{Solving the Multimessenger Puzzle of the AGN-starburst Composite Galaxy NGC 1068}},
  journal       = {\apj},
  year          = 2022,
  month         = nov,
  volume        = {939},
  number        = {1},
  eid           = {43},
  pages         = {43},
  doi           = {10.3847/1538-4357/ac9588},
  archiveprefix = {arXiv},
  eprint        = {2207.00102},
  primaryclass  = {astro-ph.HE},
  adsurl        = {https://ui.adsabs.harvard.edu/abs/2022ApJ...939...43E}
}

@article{Inoue:2019ApJ,
  author        = {{Inoue}, Yoshiyuki and {Khangulyan}, Dmitry and {Inoue}, Susumu and {Doi}, Akihiro},
  title         = {{On High-energy Particles in Accretion Disk Coronae of Supermassive Black Holes: Implications for MeV Gamma-rays and High-energy Neutrinos from AGN Cores}},
  journal       = {\apj},
  year          = 2019,
  month         = jul,
  volume        = {880},
  number        = {1},
  eid           = {40},
  pages         = {40},
  doi           = {10.3847/1538-4357/ab2715},
  archiveprefix = {arXiv},
  eprint        = {1904.00554},
  primaryclass  = {astro-ph.HE},
  adsurl        = {https://ui.adsabs.harvard.edu/abs/2019ApJ...880...40I}
}

@inbook{Murase2023HighAGN,
	author = {Murase, Kohta and Stecker, Floyd W.},
	booktitle = {The {Encyclopedia} of {Cosmology}},
	doi = {10.1142/9789811282645_0010},
	isbn = {9789811282638},
	year = {2023},
	month = {oct 10},
	pages = {483--540},
	publisher = {WORLD SCIENTIFIC},
	title = {High-{Energy} {Neutrinos} from {Active} {Galactic} {Nuclei}},
	url = {http://dx.doi.org/10.1142/9789811282645_0010},
}

@article{Murase:2020PhRvL,
  author        = {{Murase}, Kohta and {Kimura}, Shigeo S. and {M{\'e}sz{\'a}ros}, Peter},
  title         = {{Hidden Cores of Active Galactic Nuclei as the Origin of Medium-Energy Neutrinos: Critical Tests with the MeV Gamma-Ray Connection}},
  journal       = {\prl},
  year          = 2020,
  month         = jul,
  volume        = {125},
  number        = {1},
  eid           = {011101},
  pages         = {011101},
  doi           = {10.1103/PhysRevLett.125.011101},
  archiveprefix = {arXiv},
  eprint        = {1904.04226},
  primaryclass  = {astro-ph.HE},
  adsurl        = {https://ui.adsabs.harvard.edu/abs/2020PhRvL.125a1101M}
}

@article{Padovani:2017AApR,
  author        = {{Padovani}, P. and others},
  title         = {{Active galactic nuclei: what's in a name?}},
  journal       = {\aapr},
  year          = 2017,
  month         = aug,
  volume        = {25},
  number        = {1},
  eid           = {2},
  pages         = {2},
  doi           = {10.1007/s00159-017-0102-9},
  archiveprefix = {arXiv},
  eprint        = {1707.07134},
  primaryclass  = {astro-ph.GA},
  adsurl        = {https://ui.adsabs.harvard.edu/abs/2017A&ARv..25....2P}
}

@article{Blanco2025Neutrino,
	author = {Blanco, Carlos and Hooper, Dan and Linden, Tim and Pinetti, Elena},
	journal = {Physical Review D},
	doi = {10.1103/wnjh-7nwp},
	issn = {2470-0010},
	number = {12},
	year = {2025},
	month = {dec 5},
	publisher = {American Physical Society (APS)},
	title = {Neutrino and gamma-ray emissions from {NGC} 1068},
	url = {http://dx.doi.org/10.1103/wnjh-7nwp},
	volume = {112},
}

@article{4FGL,
	author = {Abdollahi, S. and others},
	journal = {\apjs},
	doi = {10.3847/1538-4365/ab6bcb},
	issn = {0067-0049},
	number = {1},
	year = {2020},
	month = {mar 1},
	pages = {33},
	publisher = {American Astronomical Society},
	title = {Fermi {Large} {Area} {Telescope} {Fourth} {Source} {Catalog}},
	url = {http://dx.doi.org/10.3847/1538-4365/ab6bcb},
	volume = {247},
}

@article{Abbasi2010Calibration,
	author = {Abbasi, R. and others},
	journal = {Nuclear Instruments and Methods in Physics Research Section A: Accelerators, Spectrometers, Detectors and Associated Equipment},
	doi = {10.1016/j.nima.2010.03.102},
	issn = {0168-9002},
	number = {1-3},
	year = {2010},
	month = {6},
	pages = {139--152},
	publisher = {Elsevier BV},
	title = {Calibration and characterization of the {IceCube} photomultiplier tube},
	url = {http://dx.doi.org/10.1016/j.nima.2010.03.102},
	volume = {618},
}

@article{Aartsen2019MeasurementNuTau,
	author = {Aartsen, M. G. and others},
	journal = {Physical Review D},
	doi = {10.1103/physrevd.99.032007},
	issn = {2470-0010},
	number = {3},
	year = {2019},
	month = {feb 15},
	publisher = {American Physical Society (APS)},
	title = {Measurement of atmospheric tau neutrino appearance with {IceCube} {DeepCore}},
	url = {http://dx.doi.org/10.1103/PhysRevD.99.032007},
	volume = {99},
}

@article{Bayer2020look,
	author = {Bayer, Adrian E. and Seljak, Uro{\v s}},
	journal = {Journal of Cosmology and Astroparticle Physics},
	doi = {10.1088/1475-7516/2020/10/009},
	issn = {1475-7516},
	number = {10},
	year = {2020},
	month = {oct 2},
	pages = {009--009},
	publisher = {IOP Publishing},
	title = {The look-elsewhere effect from a unified {Bayesian} and frequentist perspective},
	url = {http://dx.doi.org/10.1088/1475-7516/2020/10/009},
	volume = {2020},
}

@article{Braun2008Methods,
	author = {Braun, Jim and Dumm, Jon and De Palma, Francesco and Finley, Chad and Karle, Albrecht and Montaruli, Teresa},
	journal = {Astroparticle Physics},
	doi = {10.1016/j.astropartphys.2008.02.007},
	issn = {0927-6505},
	number = {4},
	year = {2008},
	month = {5},
	pages = {299--305},
	publisher = {Elsevier BV},
	title = {Methods for point source analysis in high energy neutrino telescopes},
	url = {http://dx.doi.org/10.1016/j.astropartphys.2008.02.007},
	volume = {29},
}

@article{Fisher1953Dispersion,
	author = {Fisher, R.},
	journal = {Proceedings of the Royal Society A: Mathematical, Physical and Engineering Sciences},
	doi = {10.1098/rspa.1953.0064},
	issn = {1364-5021},
	number = {1130},
	year = {1953},
	month = {may 7},
	pages = {295--305},
	publisher = {Oxford University Press (OUP)},
	title = {Dispersion on a {Sphere}},
	url = {http://dx.doi.org/10.1098/rspa.1953.0064},
	volume = {217},
}

@article{KDE:Poluektov_2015,
   title={Kernel density estimation of a multidimensional efficiency profile},
   volume={10},
   ISSN={1748-0221},
   url={http://dx.doi.org/10.1088/1748-0221/10/02/P02011},
   DOI={10.1088/1748-0221/10/02/p02011},
   number={02},
   journal={Journal of Instrumentation},
   publisher={IOP Publishing},
   author={{Poluektov}, A.},
   year={2015},
   month=feb, pages={P02011–P02011} }

@article{Icecube:2024Seyfert,
doi = {10.3847/1538-4357/addd05},
url = {https://dx.doi.org/10.3847/1538-4357/addd05},
year = {2025},
month = {jul},
publisher = {The American Astronomical Society},
volume = {988},
number = {1},
pages = {141},
author = {{Abbasi}, R. and others},
title = {IceCube Search for Neutrino Emission from X-Ray Bright Seyfert Galaxies},
journal = {\apj},
}

@article{Ackermann2013DETERMINATION,
	author = {Ackermann, M. and others},
	journal = {\apj},
	doi = {10.1088/0004-637x/765/1/54},
	issn = {0004-637X},
	number = {1},
	year = {2013},
	month = {feb 15},
	pages = {54},
	publisher = {American Astronomical Society},
	title = {DETERMINATION {OF} {THE} {POINT}-{SPREAD} {FUNCTION} {FOR} {THE}\textit{{FERMI}}{LARGE} {AREA} {TELESCOPE} {FROM} {ON}-{ORBIT} {DATA} {AND} {LIMITS} {ON} {PAIR} {HALOS} {OF} {ACTIVE} {GALACTIC} {NUCLEI}},
	url = {http://dx.doi.org/10.1088/0004-637X/765/1/54},
	volume = {765},
}

@article{Eilers1996Flexible,
	author = {Eilers, Paul H. C. and Marx, Brian D.},
	journal = {Statistical Science},
	doi = {10.1214/ss/1038425655},
	issn = {0883-4237},
	number = {2},
	year = {1996},
	month = {may 1},
	publisher = {Institute of Mathematical Statistics},
	title = {Flexible smoothing with {B}-splines and penalties},
	url = {http://dx.doi.org/10.1214/ss/1038425655},
	volume = {11},
}

@article{Gorski2005HEALPix,
	author = {Gorski, K. M. and Hivon, E. and Banday, A. J. and Wandelt, B. D. and Hansen, F. K. and Reinecke, M. and Bartelmann, M.},
	journal = {\apj},
	doi = {10.1086/427976},
	issn = {0004-637X},
	number = {2},
	year = {2005},
	month = {4},
	pages = {759--771},
	publisher = {American Astronomical Society},
	title = {HEALPix: A {Framework} for {HighResolution} {Discretization} and {Fast} {Analysis} of {Data} {Distributed} on the {Sphere}},
	url = {http://dx.doi.org/10.1086/427976},
	volume = {622},
}

@article{Sidak1967Rectangular,
	author = {{\v S}id{\' a}k, Zbyn{\v e}k},
	journal = {Journal of the American Statistical Association},
	doi = {10.1080/01621459.1967.10482935},
	issn = {0162-1459},
	number = {318},
	year = {1967},
	month = {6},
	pages = {626--633},
	publisher = {Informa UK Limited},
	title = {Rectangular {Confidence} {Regions} for the {Means} of {Multivariate} {Normal} {Distributions}},
	url = {http://dx.doi.org/10.1080/01621459.1967.10482935},
	volume = {62},
}

@article{Ackermann2011Cocoon,
	author = {Ackermann, M. and others},
	journal = {Science},
	doi = {10.1126/science.1210311},
	issn = {0036-8075},
	number = {6059},
	year = {2011},
	month = {nov 25},
	pages = {1103--1107},
	publisher = {American Association for the Advancement of Science (AAAS)},
	title = {A {Cocoon} of {Freshly} {Accelerated} {Cosmic} {Rays} {Detected} by {Fermi} in the {Cygnus} {Superbubble}},
	url = {http://dx.doi.org/10.1126/science.1210311},
	volume = {334},
}

@inproceedings{Wang2023Time,
	author = {Wang, Xubin and others},
	booktitle = {Proceedings of 38th {International} {Cosmic} {Ray} {Conference} --- {PoS}({ICRC2023})},
	doi = {10.22323/1.444.1079},
	year = {2023},
	month = {aug 17},
	pages = {1079},
	organization = {Sissa Medialab},
	title = {Time-dependent and {Time}-independent {Directional} {Search} for {High}-{Energy} {Astrophysical} {Neutrino} {Point} {Sources} in {Super}-{Kamiokande}},
	url = {http://dx.doi.org/10.22323/1.444.1079},
}

@ARTICLE{2025arXiv251107239A,
       author = {{Albert}, A. and others},
        title = "{Search for steady and flaring neutrino emission from cosmic sources using the complete ANTARES dataset}",
      journal = {arXiv e-prints},
         year = 2025,
        month = nov,
          eid = {arXiv:2511.07239},
        pages = {arXiv:2511.07239},
          doi = {10.48550/arXiv.2511.07239},
archivePrefix = {arXiv},
       eprint = {2511.07239},
 primaryClass = {astro-ph.HE},
       adsurl = {https://ui.adsabs.harvard.edu/abs/2025arXiv251107239A}
}

@article{Neronov2024Neutrino,
	author = {Neronov, A. and Savchenko, D. and Semikoz, D. V.},
	journal = {Physical Review Letters},
	doi = {10.1103/physrevlett.132.101002},
	issn = {0031-9007},
	number = {10},
	year = {2024},
	month = {mar 4},
	publisher = {American Physical Society (APS)},
	title = {Neutrino {Signal} from a {Population} of {Seyfert} {Galaxies}},
	url = {http://dx.doi.org/10.1103/PhysRevLett.132.101002},
	volume = {132},
}

@article{Abbasi2025SearchXR,
	author = {Abbasi, R. and others},
	journal = {\apj},
	doi = {10.3847/1538-4357/ada94b},
	issn = {0004-637X},
	number = {2},
	year = {2025},
	month = {mar 4},
	pages = {131},
	publisher = {American Astronomical Society},
	title = {Search for {Neutrino} {Emission} from {Hard} {X}-{Ray} {AGN} with {IceCube}},
	url = {http://dx.doi.org/10.3847/1538-4357/ada94b},
	volume = {981},
}

@article{Aiello2024Measurement,
	author = {Aiello, S. and others},
	journal = {Journal of High Energy Physics},
	doi = {10.1007/jhep10(2024)206},
	issn = {1029-8479},
	number = {10},
	year = {2024},
	month = {oct 29},
	publisher = {{Springer Science and Business Media LLC}},
	title = {Measurement of neutrino oscillation parameters with the first six detection units of {KM3NeT}/{ORCA}},
	url = {http://dx.doi.org/10.1007/JHEP10(2024)206},
	volume = {2024},
}

@article{Sgaard2023GraphNeT,
	author = {S\o{}gaard, Andreas and \O{}rs\o{}e, Rasmus F. and Holm, Morten and Bozianu, Leon and Rosted, Aske and Petersen, Troels C. and Iversen, Kaare Endrup and Hermansen, Andreas and Guggenmos, Tim and Andresen, Peter and Minh, Martin Ha and Neste, Ludwig and Holmes, Moust and Pont{\' e}n, Axel and DeHolton, Kayla Leonard and Eller, Philipp},
	journal = {Journal of Open Source Software},
	doi = {10.21105/joss.04971},
	issn = {2475-9066},
	number = {85},
	year = {2023},
	month = {may 12},
	pages = {4971},
	publisher = {The Open Journal},
	title = {GraphNeT: Graph neural networks for neutrino telescope
event reconstruction},
	url = {http://dx.doi.org/10.21105/joss.04971},
	volume = {8},
}

@ARTICLE{1962AJ.....67..471K,
       author = {{King}, Ivan},
        title = "{The structure of star clusters. I. an empirical density law}",
      journal = {\aj},
         year = 1962,
        month = oct,
       volume = {67},
        pages = {471},
          doi = {10.1086/108756},
       adsurl = {https://ui.adsabs.harvard.edu/abs/1962AJ.....67..471K}
}

@article{Abbott2008Search,
	author = {Abbott, B. and others},
	journal = {Physical Review D},
	doi = {10.1103/physrevd.77.062004},
	issn = {1550-7998},
	number = {6},
	year = {2008},
	month = {mar 17},
	publisher = {American Physical Society (APS)},
	title = {Search for gravitational waves associated with 39 gamma-ray bursts using data from the second, third, and fourth {LIGO} runs},
	url = {http://dx.doi.org/10.1103/PhysRevD.77.062004},
	volume = {77},
}

@article{Albert2019ANTARESold,
	author = {Albert, A. and others},
	journal = {\apj},
	doi = {10.3847/1538-4357/ab253c},
	issn = {0004-637X},
	number = {2},
	year = {2019},
	month = {jul 10},
	pages = {108},
	publisher = {American Astronomical Society},
	title = {ANTARES {Neutrino} {Search} for {Time} and {Space} {Correlations} with {IceCube} {High}-energy {Neutrino} {Events}},
	url = {http://dx.doi.org/10.3847/1538-4357/ab253c},
	volume = {879},
}

@book{rudin_1976_principles,
  author    = {Rudin, Walter},
  title     = {Principles of Mathematical Analysis},
  edition   = {3rd},
  publisher = {McGraw-Hill},
  year      = {1976}
}

@Inbook{Hofmann2022,
author="Hofmann, Werner
and Zanin, Roberta",
editor="Bambi, Cosimo
and Santangelo, Andrea",
title="The Cherenkov Telescope Array",
bookTitle="Handbook of X-ray and Gamma-ray Astrophysics",
year="2022",
publisher="Springer Nature Singapore",
address="Singapore",
pages="1--47",
isbn="978-981-16-4544-0",
doi="10.1007/978-981-16-4544-0_70-1",
url="https://doi.org/10.1007/978-981-16-4544-0_70-1"
}
\bibliographystyle{aasjournalv7}



\end{document}